\documentclass[review]{elsarticle}

\usepackage{lineno}
\usepackage{url}
\usepackage{float}
\usepackage{boldline}
\usepackage{booktabs}
\usepackage{amssymb,amsmath,amsthm}
\newtheorem{mydef}{Definition}

\newtheorem{mytheo}{Theorem}
\usepackage{multirow}
\usepackage{hyperref}
\hypersetup{
    colorlinks=true,
    linkcolor=magenta,
    filecolor=blue,      
    urlcolor=blue,
    citecolor=blue,
    pdftitle={Sharelatex Example},
    bookmarks=true,
    }
\usepackage{mathtools}

\usepackage[linesnumbered,ruled,vlined]{algorithm2e}
\usepackage[framemethod=TikZ]{mdframed}
\mdfdefinestyle{MyFrame}{%
	linecolor=blue,
	outerlinewidth=1.0pt,
	roundcorner=20pt,
	innertopmargin=\baselineskip,
	innerbottommargin=\baselineskip,
	innerrightmargin=20pt,
	innerleftmargin=20pt,
	backgroundcolor=gray!50!white}
\usepackage{tcolorbox}
\modulolinenumbers[5]

\journal{Journal of \LaTeX\ Templates}

\biboptions{authoryear}

\begin{document}

\begin{frontmatter}

\title{Budgeted Influence and Earned Benefit Maximization with Tags in Social Networks
\tnoteref{mytitlenote}}
\tnotetext[mytitlenote]{A part of this paper has been previously published as \cite{banerjee2020budgeted}}

\author{Suman Banerjee}
\address{Department of Computer Science and Engineering,\\
Indian Institute of Technology, Jammu\\
Jammu \& Kashmir-181221, India.\\
\ead{suman.banerjee@iitjammu.ac.in}}

\author{Bithika Pal}
\address{Department of Computer Science and Engineering,\\
Indian Institute of Technology, Kharagpur, \\
West Bengal-721302, India.\\
\ead{bithikapal@iitkgp.ac.in}}

%


\begin{abstract}
Given a social network, where each user is associated with a selection cost, the problem of \textsc{Budgeted Influence Maximization} (\emph{BIM Problem} in short) asks to choose a subset of them (known as seed users) within an allocated budget whose initial activation leads to the maximum number of influenced nodes. Existing Studies on this problem do not consider the tag-specific influence probability. However, in reality, influence probability between two users always depends upon the context (e.g., sports, politics, etc.). To address this issue, in this paper we introduce the \textsc{Tag\mbox{-}Based Budgeted Influence Maximization problem} (\emph{TBIM Problem} in short), where along with the other inputs, a tag set (each of them is also associated with a selection cost) is given, each edge of the network has the tag specific influence probability, and here the goal is to select influential users as well as influential tags within the allocated budget to maximize the influence. Considering the fact that real-world campaigns targeted in nature, we also study the \textsc{Earned Benefit Maximization} Problem in tag specific influence probability setting, which formally we call the \textsc{Tag\mbox{-}Based Earned Benefit Maximization problem} (\emph{TEBM Problem} in short). For this problem along with the inputs of the TBIM Problem, we are given a subset of the nodes as target users, and each one of them is associated with a benefit value that can be earned by influencing them. Considering the fact that different tag has different popularity across the communities of the same network, we propose three methodologies that work based on \emph{effective marginal influence gain computation}. The proposed methodologies have been analyzed for their time and space requirements. We evaluate the methodologies with three publicly available social network datasets, and observe, that these can select seed nodes and influential tags, which leads to more number of influenced nodes and more amount of earned benefit compared to the baseline methods.
\end{abstract}


\end{frontmatter}

\section{Introduction}
A social network is an interconnected structure among a group of agents formed for social interactions \cite{carrington2005models}. One of the important phenomena of social networks is the diffusion of information. Based on the diffusion process, a well-studied problem in the domain of computational social network analysis is the \emph{Social Influence Maximization} (\emph{SIM Problem}), which has an immediate application in the context of \emph{viral marketing}. The goal here is to get wider publicity for a product by initially distributing a limited number of free samples to highly influential users. For a given social network and a positive integer $k$, the SIM Problem asks to select $k$ users for initial activation to maximize the influence in the network. Due to potential application across multiple domains such as personalized recommendation \cite{zhang2013socconnect}, feed ranking \cite{ienco2010meme}, viral marketing \cite{chen2010scalable}, this problem remains the central theme of research since last one and half decades or so. Please look into \cite{li2018influence, peng2018influence,banerjee2020survey} for recent surveys.
\par Recently, a variant of this problem has been introduced by Nguyen and Zheng \cite{nguyen2013budgeted}, where the users of the network are associated with a selection cost and the seed set selection is to be done within an allocated budget to maximize the influence in the network. There are a few solution methodologies of the problem such as directed acyclic graph (DAG)\mbox{-}based heuristics and $(1-\frac{1}{\sqrt{e}})$\mbox{-}factor approximation algorithm by \cite{nguyen2013budgeted}, balanced seed selection heuristics by \cite{han2014balanced}, integer programming\mbox{-}based approach \cite{guney2017optimal}, community\mbox{-}based  solution approach by \cite{banerjee2019combim}. In all these studies it is implicitly assumed that irrespective of the context, influence probability between two users will be the same, i.e., there is a single influence probability associated with every edge. However, in reality, the scenario is different. It is natural that a sportsman can influence his friend in any sports related news with more probability compared to political news. This means the influence probability between any two users is context specific, and hence, in Twitter, a follower will re\mbox{-}tweet if the tweet contains some specific hash tags. To address this issue, in this paper we introduce the \emph{Tag\mbox{-}based Budgeted Influence Maximization Problem (TBIM Problem)}, which considers the tag specific influence probability assigned with every edge of the network. 
\par Consider the case of a commercial campaign, where an E\mbox{-}Commerce house promotes its newly developed product. It is quite natural  that not all the customers present in their database will be interested in the product. Hence, it is not really beneficial to advertise a product to a user, if he is not at all interested and this can be easy to predict from his browsing/liking/rating history. This implies that the advertisement needs to be conducted among the group of target users. It is quite natural that the target users are associated with a benefit value that can be earned by the E\mbox{-}Commerce house by influencing the corresponding user. Hence, in a targeted advertisement setting the natural problem is that given a social network where every user is associated with a selection cost, a set of target users with a benefit value. The goal here is to choose a seed set within the allocated budget to maximize the earned benefit by influencing the target users. Formally, this problem has been referred to as the \emph{Earned Benefit Maximization} Problem \cite{banerjee2020earned}. There are several studies by Banerjee et al. \cite{banerjee2020earned,banerjee2019maximizing, banerjee2019maximizingw} and also by several other authors as well in this direction \cite{tang2017profit, zhu2017maximizing, zhou2019cost}. However, none of these studies considers the tag specific influence probability, though it is a practical concern as mentioned previously. In this paper, we study this problem in tag specific influence probability setting and formally call as \textsc{Tag-Based Earned Benefit Maximization} Problem (\emph{TEBM Problem}).
\par
\begin{figure}[t]
  \centering
  \includegraphics[width=\linewidth]{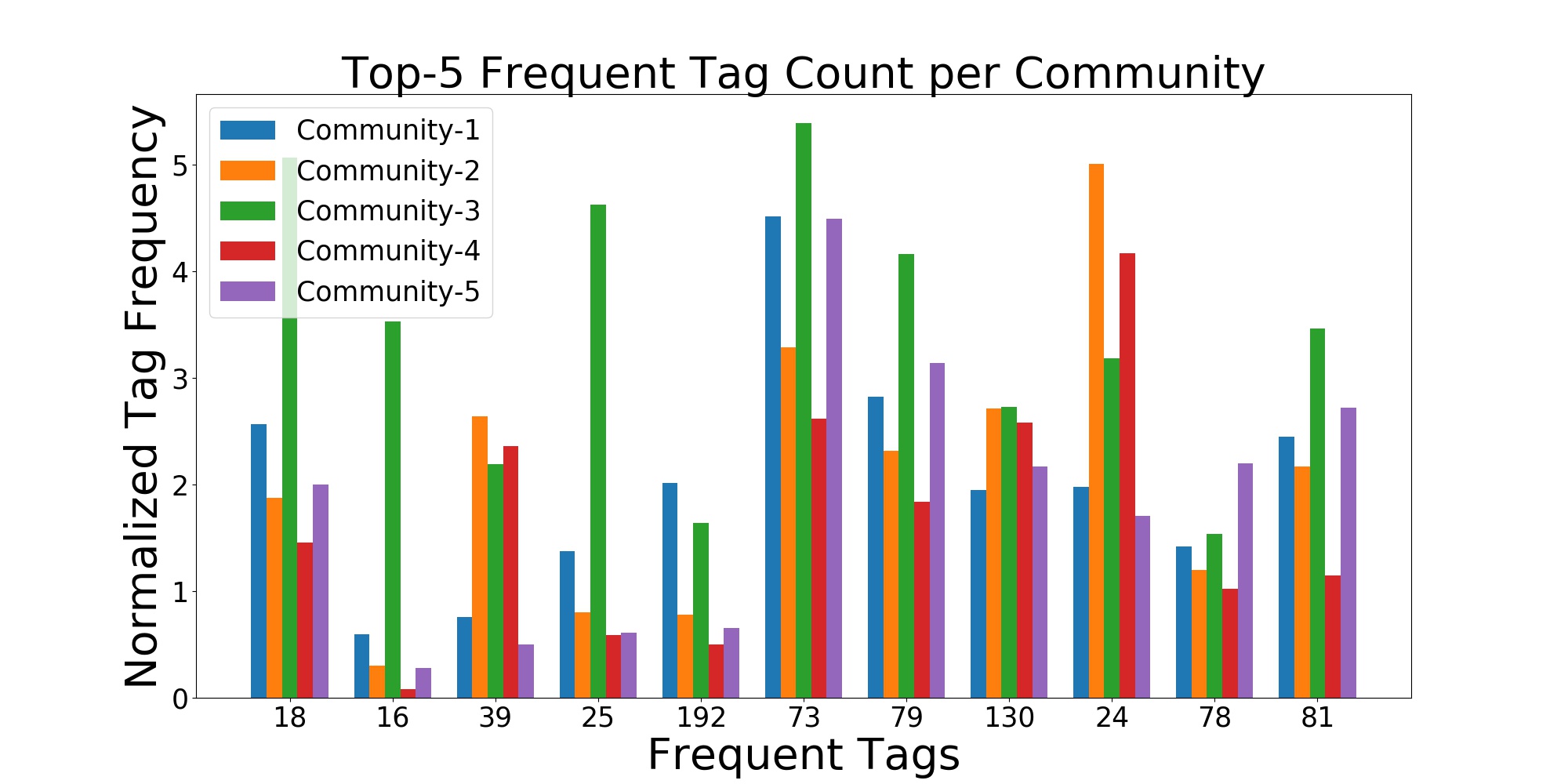}
  \caption{Community specific distribution top 5 tags in `Last.fm' dataset}
  \label{Fig:1}
\end{figure}

 Recently, \cite{ke2018finding} studied the problem of finding $k$ seed nodes and $r$ influential tags in a given social network. However,  their study has two drawbacks. First, in reality, most of the social networks are formed by rational human beings. Hence, once a node is selected as a seed then incentivization is required (e.g., free or discounted sample of the item to be advertised). In practical cases, E-Commerce houses hire some online social media platforms to run the commercial campaign. The message for this purpose may be composed of many animations, videos, images, and also texts (together referred to as tags). To display a tag in an on\mbox{-}line advertisement also associated with some cost. Hence, online social media platforms will charge money based on the tags present in the message. So, the tags that have been used to construct the message will also be associated with a selection cost. Their study does not consider these issues. Secondly, in \cite{ke2018finding}'s study, they have done the tag selection process at the network level. However, in reality, popular tags may vary from one community to another in the same network. Figure \ref{Fig:1} shows community-wise distribution of Top $5$ tags for the Last.fm dataset. From the figure, it is observed that Tag No. $16$ has the highest popularity in Community $3$. However, its popularity is very less in Community $4$. This signifies that tag selection at the network level may not be always helpful to spread influence in each community of the network. To mitigate these issues, in this paper, we propose three solution methodologies for the TBIM Problem, where the tag selection is done community-wise. To summarize, the main contributions of this paper are as follows:
\begin{itemize}
\item Considering the tag specific influence probability, in this paper, we introduce the \textsc{Tag\mbox{-}Based Budgeted Influence Maximization} Problem (TBIM Problem) and \textsc{Tag\mbox{-}Based Earned Benefit Maximization} Problem.
\item For both these problems, we propose two different marginal influence gain-based approaches with their detailed analysis.
\item To increase the scalability of the proposed methodologies, we propose an efficient pruning technique.
\item The proposed methodologies have been implemented with three publicly available datasets and a set of experiments have been performed to show their effectiveness. 
\end{itemize}
\par The rest of the paper is arranged as follows: Section \ref{Sec:RW} discusses some relevant studies from the literature. Section \ref{Sec:BPD} contains some background material and defines the TBIM Problem formally. The proposed solution methodologies for this problem have been described in Section \ref{Sec:PM}. In Section \ref{Sec:ER}, we report the experimental results of the proposed methodologies. Finally, Section \ref{Sec:CFD} concludes of our study.
\section{Related Work} \label{Sec:RW}
Our study is related to the Social and Budgeted Influence maximization, Keyword\mbox{-}Based Influence Maximization and Community\mbox{-}Based Solution Methodologies for the Influence Maximization. Here we describe them one by one.
\paragraph{\textbf{Social/Budgeted Influence maximization}} Given a social network and positive integer $k$, the problem of Social Influence Maximization asked to select Top\mbox{-}k influential users in the network for initial activation to maximize the influence. This problem has been initially proposed by \cite{domingos2001mining} in the context of viral marketing. Later, \cite{kempe2003maximizing} resolves the complexity issues of this problem and proposed an incremental greedy strategy that admits $(1-\frac{1}{e})$\mbox{-} factor approximation guarantee. This study triggers extensive research work and hence, a vast amount of literature is available. Most of the solution methodologies are concerned either about the quality of the selected seed set (i.e., the optimality gap) or about the scalability issues (i.e., when the problem size increases how the computational time changes). Proposed solution methodologies can be classified into different categories such as \emph{approximation algorithms} (e.g., CELF \cite{leskovec2007cost}, CELF++ \cite{goyal2011celf}, MIA and PMIA \cite{wang2012scalable}), sketch\mbox{-}based approaches (e.g., TIM \cite{tang2014influence}, IMM \cite{tang2015influence}), heuristic approaches (e.g., SPIN \cite{narayanam2011shapley}, ASIM \cite{galhotra2015asim} and many more), soft computing\mbox{-}based approaches (GA \cite{bucur2016influence} \cite{zhang2017maximizing}, PSO  \cite{gong2016influence}) and so on.
\par Recently, a variant of the SIM Problem has been introduced by  \cite{nguyen2013budgeted}, where the nodes of the network are associated with a selection cost and the goal is to select a subset of the nodes within an allocated budget for initial activation to maximize the influence. This Problem is known as the Budgeted Influence Maximization Problem. Available solution methodologies for this problems are as follows.  \cite{nguyen2013budgeted} proposed a $(1-\frac{1}{\sqrt{e}})$\mbox{-}factor approximation algorithm and two DAG\mbox{-}based heuristics for the BIM Problem. \cite{han2014balanced} proposed a set of heuristics considering both influential nodes and also cost effective nodes.  \cite{guney2017optimal} proposed an integer programming\mbox{-}based approach under the independent cascade model of diffusion. Recently,  \cite{banerjee2019combim} proposed a community\mbox{-}based solution approach for this problem. However, none of these studies consider the tag-specific influence probability. 
\paragraph{\textbf{Tag/Keyword\mbox{-}Based Influence Maximization}} As mentioned previously, keywords are important in the case of online advertising. There are a few studies available, where keyword-based influence probabilities have been taken into consideration. \cite{li2015real} studied the SIM Problem, where the goal is to select a subset of the users (known as target users). Recently, \cite{ke2018finding} studied the problem for finding the seed nodes and initial tag set jointly and they proposed heuristic solutions for this problem. \cite{fan2018octopus} developed an online topic-based influence analysis system that discovers keyword\mbox{-}based influential users in the network as one of its tasks with two others. However, none of these studies consider the selection cost of the tags, which is a practical concern as mentioned earlier.
\paragraph{\textbf{Community\mbox{-}based Influence Maximization}} As the real\mbox{-}world social networks are gigantic in size and pose a strong community structure, recently several methods have been proposed where the community structure of the network has been exploited. In this direction, the first study was made by \cite{wang2010community}  \cite{chen2012exploring,chen2014cim}, and their method outperforms many existing methods in terms of scalability and efficiency.  \cite{bozorgi2016incim} proposed a community\mbox{-}based approach under the linear threshold diffusion model, where the goal is to find influential communities by considering the global influence of the users.  \cite{shang2017cofim} proposed a community\mbox{-}based framework that leads to acceptable computational time for large scale networks.  \cite{bozorgi2017community} proposed a solution methodology using the community structure of the network under the competitive linear threshold model. \cite{hosseini2017community} proposed  a community\mbox{-}based approach that shows a balance between between  effectiveness and efficiency. Recently, there are several studies in this direction \cite{huang2019community,li2018community,singh2019c2im}.
\paragraph{\textbf{Benefit/Profit Maximization using Social Networks}} In recent times there are several studies that focus on benefit/profit maximization in online social networks \cite{tong2018coupon,tang2017profit,huang2020efficient}. \cite{tong2018coupon} considered the coupon allocation problem in profit maximization. By utilizing the double greedy algorithm proposed by \cite{buchbinder2015tight}, they showed that their proposed approaches can achieve $\frac{1}{2}$-factor approximation guarantee with high probability. Later, \cite{liu2020profit} also studied this problem under the  \emph{independent cascade model with coupon and valuation} and proposed  the PMCA Algorithm based on a local search technique proposed in \cite{feige2011maximizing}. If the submodular profit function is non-negative for every subset of the users, then PMCA can achieve an approximation ratio $\frac{1}{3}$. \cite{tang2017profit} studied the profit maximization problem and proposed two different approaches for solving this problem; namely an incremental greedy hill-climbing approach, and a double greedy algorithm. 
\par In this paper, we propose three solution methodologies for the TBIM and TEBM Problem, which exploits the community structure of the network.
\section{Background and Problem Definition} \label{Sec:BPD}
We assume that the input social network is represented as a directed and edge-weighted graph $\mathcal{G}(\mathcal{V}, \mathcal{E}, \mathcal{P})$, where the vertex set, $\mathcal{V}(\mathcal{G})=\{u_1, u_2, \ldots, u_n\}$ is the set of $n$ users, the edge set $\mathcal{E}(\mathcal{G})=\{e_1, e_2, \ldots, e_m\}$ is the set of $m$ social ties among the users. Along with $\mathcal{G}$, we are also given with a tag set $T=\{t_1, t_2, \ldots, t_a\}$ relevant to the users of the network. Here, $\mathcal{P}$ is the edge weight function that assigns each edge to its tag-specific influence probability, i.e., $\mathcal{P}: \mathcal{E}(\mathcal{G}) \longrightarrow (0,1]^{|T|}$. This means that each edge of the network is associated with an influence probability vector, whose each dimension is for a particular tag. For all $(u_iu_j) \in \mathcal{E}(\mathcal{G})$, we denote its corresponding influence probability vector as $\mathcal{P}_{u_i \rightarrow u_j}$. Also, for a particular tag $t \in T$ and an edge $(u_iu_j) \in \mathcal{E}(\mathcal{G})$, we denote the influence probability of the edge $(u_iu_j)$ for the tag $t$ as $\mathcal{P}_{u_i \rightarrow u_j}^{t}$. This can be interpreted as the conditional probability of the edge $(u_iu_j)$ given the tag $t$, i.e., $\mathcal{P}_{u_i \rightarrow u_j}^{t}=Pr((u_iu_j)|t)$. Now, a subset of the available tags $T^{'} \subseteq T$ which are relevant to the campaign may be used. So, it is important how to compute the effective probability for each edge and this depends upon how the selected tags are aggregated. In this study, we perform the \emph{independent tag aggregation}, which is defined as follows. 
\begin{mydef}[Independent Tag Aggregation]
Given a social network $\mathcal{G}(\mathcal{V}, \mathcal{E}, \mathcal{P})$, the tag set $T^{'} \subseteq T$, the aggregated influence probability of the  edge $(u_iu_j)$ can be computed as follows:
\begin{equation}\label{Eq:1}
\mathcal{P}_{u_i \rightarrow u_j}^{T^{'}}=1- \underset{t \in T^{'}}{\prod} 
(1-\mathcal{P}_{u_i \rightarrow u_j}^{t})
\end{equation} 
\end{mydef}
\par Now, to conduct a campaign using a social network, a subset of the users need to be selected initially as seed nodes. The number of influenced nodes will depend upon which influence propagation model we are using. In our study, we assume that the influence is propagated based on the \emph{Maximum Influence Arborance (MIA)} diffusion model, which is discussed next. 
\paragraph{Diffusion in Social Networks} We denote the seed set by $\mathcal{S}$. The users in the set $\mathcal{S}$  are informed initially, and the others are ignorant about the information. These seed users start the diffusion, and the information is diffused by the rule of an information diffusion model. There are many such rules proposed in the literature. One of them is the MIA Model \citep{wang2012scalable}. Recently, this model has been used by many existing studies in influence maximization \citep{chen2011influence}. In our study also, we assume that the information is diffused by the rule of MIA Model. Before defining the MIA Model, we first state a few preliminary definitions. 
\begin{mydef}[Propagation Probability of a Path]
Given two vertices $u_i, u_j \in V(\mathcal{G})$, let $\mathbb{P}(u_i,u_j)$  denotes the set of paths from the vertex $u_i$ to $u_j$. For any arbitrary path $p \in \mathbb{P}(u_i,u_j)$ the propagation probability  is defined as the product of the influence probabilities of the constituent edges of the path.
\begin{equation}
  \mathcal{P}(p) =
  \begin{cases}
    \underset{(u_iu_j) \in \mathcal{E}(p)}{\prod}  \mathcal{P}^{T^{'}}_{u_i \rightarrow u_j} & \text{if $\mathbb{P}(u_i,u_j) \neq \phi$} \\
    0 & \text{otherwise}\\
  \end{cases}
\end{equation}
Here, $\mathcal{E}(p)$ denotes the edges that constitute the path $p$.
\end{mydef}
\begin{mydef}[Maximum Probabilistic Path]
Given two vertices $u_i,u_j \in \mathcal{V}(\mathcal{G})$, the maximum probabilistic path is the path with the maximum propagation probability and denoted as $p_{(u_iu_j)}^{max}$. Hence,
\begin{equation}
p_{(u_iu_j)}^{max}=\underset{p \in \mathbb{P}(u_i,u_j)}{argmax} \ \mathcal{P}(p)
\end{equation}
\end{mydef}
\begin{mydef}[Maximum Influence In Arborence]
For a given threshold $\theta$, the maximum influence in\mbox{-}arborence of a node $v \in \mathcal{V}(\mathcal{G})$ is defined as 
\begin{equation}
MIIA(v,\theta)= \underset{u \in \mathcal{V}(\mathcal{G}),\mathcal{P}(p_{(uv)}^{max}) \geq \theta}{\bigcup} p_{(uv)}^{max}
\end{equation}
\end{mydef}
\par Given a seed set $\mathcal{S}$ and a node $v \notin \mathcal{S}$, in MIA Model, the influence from $\mathcal{S}$ to $v$ is approximated by the rule that for any $u \in \mathcal{S}$ can influence $v$ through the paths in $p_{(uv)}^{max}$. The influence probability of a node $u \in MIIA(v, \theta)$ is denoted as $ap(u, \mathcal{S}, MIIA(v, \theta))$, which is the probability that the node $u$ will be influenced by the nodes in $\mathcal{S}$  and influence is propagated through the paths in $MIIA(v, \theta)$. This can be computed by Algorithm 2, given in \citep{wang2012scalable}. Hence, the  influence spread obtained by the seed set $\mathcal{S}$ is given by Equation \ref{Eq:Inf}.
\begin{equation} \label{Eq:Inf}
\sigma(\mathcal{S})= \underset{v \in \mathcal{V}(\mathcal{G})}{\sum} ap(v, \mathcal{S}, MIIA(v, \theta))
\end{equation} 
\paragraph{Problem Definition} To study the BIM Problem along with the input social network, we are given with the selection costs of the users which is characterized by the cost function $\mathcal{C}:\mathcal{V}(\mathcal{G}) \rightarrow \mathbb{R}^{+}$, and a fixed budget $\mathcal{B}$. For any user $u \in \mathcal{V}(\mathcal{G})$, its selection cost is denoted as $\mathcal{C}(u)$. Now, we define the BIM Problem formally.
\begin{mydef}[BIM Problem]
Given a social network $\mathcal{G}(\mathcal{V}, \mathcal{E}, \mathcal{P})$, the cost function $\mathcal{C}: \mathcal{V}(\mathcal{G}) \rightarrow \mathbb{R}^{+}$, and an allocated budget $\mathcal{B}$, the BIM Problem asks to select a subset of the nodes $\mathcal{S}$ for initial activation to maximize $\sigma(\mathcal{S})$ such that $\underset{u \in \mathcal{S}}{\sum}\mathcal{C}(u) \leq \mathcal{B}$.
\end{mydef}
Now, as mentioned previously, tags are important in any kind of campaign also popularity of tags varies across the communities of a network. Assume that the input social network has $p$ number of communities, i.e., the community set is  $\mathcal{K}=\{\mathcal{K}_{1}, \mathcal{K}_{2}, \ldots, \mathcal{K}_{\ell}\}$. Naturally, all the tags that are considered in a specific context (i.e., $T$) may not be relevant to each of the communities. We denote the relevant tags of the community $\mathcal{K}_{i}$ as $T_{\mathcal{K}_{i}}$.  It is important to observe that displaying a tag in any online platform may associate some cost, which can be characterized by the tag cost function $\mathcal{C}^{T}:T \rightarrow \mathbb{R}^{+}$. Now, for a set of given tags and seed nodes what will be the number of influenced nodes in the network? This can be defined as the tag\mbox{-}based influence function. For a given seed set $\mathcal{S}$ and tag set $T^{'}$, the tag\mbox{-}based influence function $\sigma(\mathcal{S}, T^{'})$ returns the number of influenced nodes, which is defined next. 
\begin{mydef}[Tag\mbox{-}based Influence Function] \label{Def:Tag_Inf}
Given a social network $\mathcal{G}(\mathcal{V}, \mathcal{E}, \mathcal{P})$, a seed set $\mathcal{S} \subseteq \mathcal{V}(\mathcal{G})$, tag set $T^{'} \subseteq T$, the tag\mbox{-}based influence function $\sigma^{\mathcal{T}}$ that maps each combination of subset of the nodes and tags to the number of influenced nodes, i.e., $\sigma^{\mathcal{T}}: 2^{\mathcal{V}(\mathcal{G})} \times 2^{T} \longrightarrow \mathbb{R}_{0}$.
\end{mydef}
In this paper, we introduce the \textsc{Tag\mbox{-}based Budgeted Influence Maximization} Problem (TBIM Problem) and \textsc{Tag\mbox{-}based Earned Benefit  Maximization} Problem (TEBM Problem), which we define next.
\begin{mydef}[TBIM Problem]
Given a social network $\mathcal{G}(\mathcal{V}, \mathcal{E}, \mathcal{P})$, Tag set $T$, seed cost function $\mathcal{C}^{S}: \mathcal{V}(\mathcal{G}) \rightarrow \mathbb{R}^{+}$, tag cost function $\mathcal{C}^{T}: T \rightarrow \mathbb{R}^{+}$ and the budget $\mathcal{B}$, the TBIM Problem asks to select a subset of the tags from the communities, i.e., $T_{\mathcal{K}_{i}}^{'} \subseteq T_{\mathcal{K}_{i}}$, $\forall i \in [\ell]$ (here, $ T_{\mathcal{K}_{i}}^{'}  \cap T_{\mathcal{K}_{j}}^{'}= \emptyset, \ \forall i\neq j \text{ and } T^{'}= \underset{i \in [l]}{\bigcup}T_{\mathcal{K}_{i}}$), and nodes $\mathcal{S} \subseteq \mathcal{V}(\mathcal{G})$ to maximize $\sigma^{\mathcal{T}}(\mathcal{S}, T^{'})$  such that $\underset{u \in \mathcal{S}}{\sum} \mathcal{C}^{S}(u) + \underset{i \in [\ell]}{\sum} \  \underset{t \in T_{\mathcal{K}_{i}}}{\sum} \mathcal{C}^{T}(t) \leq \mathcal{B}$.
\end{mydef}
\par Mathematically, the TBIM Problem can be posed as follows:
\begin{equation}
(\mathcal{S}^{*}, T^{*})=\underset{(\mathcal{S}, T^{'}); \\ \mathcal{S} \subseteq \mathcal{V}(\mathcal{G}), T^{'} \subseteq T; \mathcal{C}^{S}(\mathcal{S})+\mathcal{C}^{T}(T^{'}) \leq \mathcal{B}}{argmax} \ \sigma^{\mathcal{T}}(\mathcal{S}, T^{'})
\end{equation}
From the algorithmic point of view the TBIM Problem can be posed as follows:
\begin{tcolorbox}
\underline{\textbf{TBIM Problem} $(\mathcal{G}, T,  \mathcal{C}^{S}, \mathcal{C}^{T}, \mathcal{B})$}\\
\textbf{Input:} Social Network $\mathcal{G}(\mathcal{V}, \mathcal{E}, \mathcal{P})$, Tag set $T$, Seed Cost Function $\mathcal{C}^{S}$, Tag Cost Function $\mathcal{C}^{T}$, Budget $\mathcal{B}$
\vspace{0.2 cm}\\
\textbf{Problem:} Select a seed set $\mathcal{S} \subseteq \mathcal{V} (\mathcal{G})$ and tags from the communities, i.e., $T_{\mathcal{K}_{i}}^{'} \subseteq T_{\mathcal{K}_{i}}$, $\forall i \in [\ell]$ ($ T_{\mathcal{K}_{i}}^{'}  \cap T_{\mathcal{K}_{j}}^{'}= \emptyset, \ \forall i\neq j \text{ and } T^{'}= \underset{i \in [\ell]}{\bigcup}T_{\mathcal{K}_{i}}$), and nodes $\mathcal{S} \subseteq \mathcal{V}(\mathcal{G})$ to maximize $\sigma^{\mathcal{T}}(\mathcal{S}, T^{'})$  such that $\underset{u \in \mathcal{S}}{\sum} \mathcal{C}^{S}(u) + \underset{i \in [\ell]}{\sum} \  \underset{t \in T_{\mathcal{K}_{i}}}{\sum} \mathcal{C}^{T}(t) \leq \mathcal{B}$.
\vspace{0.2 cm}\\
\textbf{Output:} Seed set $\mathcal{S} \subseteq \mathcal{V} (\mathcal{G})$, and Tag set $T^{'} \subseteq T$ 
\end{tcolorbox}
As mentioned previously, in many real\mbox{-}world campaigns, the goal is to not just to maximize the influence, but also to maximize the total earned benefit  by influencing the target users. Assume that $\mathcal{D} \subseteq \mathcal{V} (\mathcal{G})$ is the set of target users and $b: \mathcal{D} \longrightarrow \mathbb{R}^{+}$ is the benefit function that assigns each target user to its benefit that can be earned by influencing the corresponding target user, $\mathcal{C}:\mathcal{V}(\mathcal{G}) \longrightarrow \mathbb{R}^{+}$ is the cost function of the users, and $\mathcal{B}$ is the allocated budget for the seed set selection. Let, $b(u)$ denotes the associated benefit with the user $u$. Now, we define the earned benefit by a seed set.

\begin{mydef}[Earned Benefit by a Seed Set]
Given a seed set $\mathcal{S}$, its earned benefit is denoted as $\beta(\mathcal{S})$ and defined as
\begin{equation}
    \beta(\mathcal{S})= \underset{u \in \mathcal{D} \cap I(\mathcal{S})}{\sum} b(u).
\end{equation}
Here, $\beta(.)$ is the earned benefit function that assigns each subset to its earned benefit that assigns each subset of its user to its earned benefit; i.e.; $\beta: 2^{\mathcal{V}(\mathcal{G})} \longrightarrow \mathbb{R}^{+}$ with $\beta(\emptyset)=0$. 
\end{mydef}
As in this study, we are studying the TEBM Problem in tag specific influence probability setting, it is worthwhile to consider the \emph{Tag-Based Earned Benefit Function}, which can be defined equivalently as the tag-based social influence function is defined in Definition \ref{Def:Tag_Inf}. Now, the goal of the \textsc{Tag\mbox{-}based Earned Benefit Maximization} Problem is to choose influential users and tags to maximize the earned benefit stated in Definition \ref{Def:TEBM}.
\begin{mydef}[\textsc{Tag-Based Earned Benefit Maximization Problem}] \label{Def:TEBM}
Given a social network $\mathcal{G}(\mathcal{V}, \mathcal{E}, \mathcal{P})$, target user seed $\mathcal{D}$, benefit function $b:\mathcal{D} \longrightarrow \mathbb{R}^{+}$, Tag set $T$, seed cost function $\mathcal{C}^{S}: \mathcal{V}(\mathcal{G}) \longrightarrow \mathbb{R}^{+}$, tag cost function $\mathcal{C}^{T}: T \longrightarrow \mathbb{R}^{+}$ and the budget $\mathcal{B}$, the TEBM Problem asks to select a subset of the tags from the communities, i.e., $T_{\mathcal{K}_{i}}^{'} \subseteq T_{\mathcal{K}_{i}}$, $\forall i \in [\ell]$ (here, $ T_{\mathcal{K}_{i}}^{'}  \cap T_{\mathcal{K}_{j}}^{'}= \emptyset, \ \forall i\neq j \text{ and } T^{'}= \underset{i \in [\ell]}{\bigcup}T_{\mathcal{K}_{i}}$), and nodes $\mathcal{S} \subseteq \mathcal{V}(\mathcal{G})$ to maximize $\beta^{\mathcal{T}}(\mathcal{S}, T^{'})$  such that $\underset{u \in \mathcal{S}}{\sum} \mathcal{C}^{S}(u) + \underset{i \in [\ell]}{\sum} \  \underset{t \in T_{\mathcal{K}_{i}}}{\sum} \mathcal{C}^{T}(t) \leq \mathcal{B}$.
\end{mydef}
\par Mathematically, the TEBM Problem can be posed as follows:
\begin{equation}
(\mathcal{S}^{*}, T^{*})=\underset{(\mathcal{S}, T^{'}); \\ \mathcal{S} \subseteq \mathcal{V}(\mathcal{G}), T^{'} \subseteq T; \mathcal{C}^{S}(\mathcal{S})+\mathcal{C}^{T}(T^{'}) \leq \mathcal{B}}{argmax} \ \beta^{\mathcal{T}}(\mathcal{S}, T^{'})
\end{equation}
From the computational point of view, the TEBM Problem can be defined as follows:

\begin{tcolorbox}
\underline{\textbf{TEBM Problem} $(\mathcal{G}, \mathcal{D}, b, T,  \mathcal{C}^{S}, \mathcal{C}^{T}, \mathcal{B})$}\\
\textbf{Input:} Social Network $\mathcal{G}(\mathcal{V}, \mathcal{E}, \mathcal{P})$, Target User Set $\mathcal{D}$, Benefit Function $b$,  Tag set $T$, Seed Cost Function $\mathcal{C}^{S}$, Tag Cost Function $\mathcal{C}^{T}$, Budget $\mathcal{B}$
\vspace{0.2 cm}\\
\textbf{Problem:} Select a seed set $\mathcal{S} \subseteq \mathcal{V} (\mathcal{G})$ and tags from the communities, i.e., $T_{\mathcal{K}_{i}}^{'} \subseteq T_{\mathcal{K}_{i}}$, $\forall i \in [\ell]$ ($ T_{\mathcal{K}_{i}}^{'}  \cap T_{\mathcal{K}_{j}}^{'}= \emptyset, \ \forall i\neq j \text{ and } T^{'}= \underset{i \in [\ell]}{\bigcup}T_{\mathcal{K}_{i}}$), and nodes $\mathcal{S} \subseteq \mathcal{V}(\mathcal{G})$ to maximize $\beta^{\mathcal{T}}(\mathcal{S}, T^{'})$  such that $\underset{u \in \mathcal{S}}{\sum} \mathcal{C}^{S}(u) + \underset{i \in [\ell]}{\sum} \  \underset{t \in T_{\mathcal{K}_{i}}}{\sum} \mathcal{C}^{T}(t) \leq \mathcal{B}$.
\vspace{0.2 cm}\\
\textbf{Output:} Seed set $\mathcal{S} \subseteq \mathcal{V} (\mathcal{G})$, and Tag set $T^{'} \subseteq T$ 
\end{tcolorbox}
Symbols that have been used in this paper is shown in Table \ref{tab:Symbols}. 
\begin{table}
  \caption{Symbols with their interpretation}
  \label{tab:Symbols}
  \begin{tabular}{||c|c||}
    \toprule
    Symbol & Interpretation \\
    \midrule
    $\mathcal{G}(\mathcal{V}, \mathcal{E}, \mathcal{P})$ & The input social network \\
    $\mathcal{V}(\mathcal{G}), \mathcal{E}(\mathcal{G})$ & The vertex set and edge set of $\mathcal{G}$ \\
    $\mathcal{D}$ & The set of target users\\
    $T$ & The tag set \\
   $\mathcal{P}$ & The edge weight function $\mathcal{P}: \mathcal{E}(\mathcal{G}) \times T \longrightarrow (0,1]^{|T|}$\\
   $\mathcal{P}_{u_i \rightarrow u_j}$ & The influence probability vector from the user $u_i$ to $u_j$\\
   $\mathcal{P}_{u_i \rightarrow u_j}^{t}$ & The influence probability from the user $u_i$ to $u_j$ for the tag $t$\\
   $\mathcal{C}^{S}$, $\mathcal{C}^{T}$  & The cost function for the users and tags \\
   $\mathcal{C}^{S}(u)$, $\mathcal{C}^{T}(t)$  & The selection cost for the user $u$, and tag $t$, respectively \\
   $b$ & Benefit function for target user \\
   $b(u)$ & Benefit of the user $u$ \\
   $\beta$ & Earned benefit function \\
   $\beta(\mathcal{S})$ & Earned benefit by the seed set $\mathcal{S}$ \\
   $\mathcal{B}$ & Allocated budget for the seed nodes and tag selection  \\
   $\mathcal{K}$ &  Set of communities of $\mathcal{G}$ \\
   $\ell$ &  The number of communities of $\mathcal{G}$, i.e., $|\mathcal{K}|=\ell$ \\
   $\pi_{i}$ & Priority of the $i$-th community \\
   $\mathbb{P}(u_i,u_j)$ &  Set of paths between  $u_i$, and $u_j$ \\
   $\mathbb{R}^{+}$ &  Set of positive real number \\ 
   $\mathbb{R}_{0}$ &  Set of positive real number including $0$ \\
   $\sigma(.)$ &  The social influence function \\
   $\sigma^{\mathcal{T}}(.)$ &  The tag\mbox{-}based social influence function  \\ 
   $\delta_{v}$, $\delta_{t}$  &  Effective Marginal Influence Gain for the user $u$ and tag $t$\\ 
   $\delta_{(v,t)}$  &  Effective Marginal Influence Gain for the user\mbox{-}tag pair $(v,t)$\\ 
   \bottomrule
\end{tabular}
\end{table}

\section{Proposed Methodologies} \label{Sec:PM}
In this section, we propose two different approaches and one subsequent improvement to select tags and seed users for initiating the diffusion process. Before stating the proposed solution approaches, we first define the \emph{Effective Marginal Influence Gain}.
\begin{mydef}[Effective Marginal Influence Gain]
Given a seed set $\mathcal{S} \subset \mathcal{V}(\mathcal{G})$, tag set $T^{'} \subset T$, the effective marginal influence gain (EMIG, henceforth) of the node $v \in  \mathcal{V}(\mathcal{G}) \setminus \mathcal{S}$ (denoted as $\delta_{v}$) with respect to the seed set $\mathcal{S}$ and tag set $T^{'}$ is defined as the ratio between the marginal influence gain to its selection cost, i.e.,
\begin{equation} \label{Eq:Node}
\delta_{v}=\frac{\sigma^{\mathcal{T}}(\mathcal{S} \cup \{v\}, T^{'})- \sigma^{\mathcal{T}}(\mathcal{S}, T^{'})}{\mathcal{C}^{S}(v)}.
\end{equation} 
In the similar way, for any tag $t \in T \setminus T^{'}$, its EMIG is defied as 
\begin{equation} \label{Eq:Tag}
\delta_{t}=\frac{\sigma^{\mathcal{T}}(\mathcal{S}, T^{'} \cup \{t\})- \sigma^{\mathcal{T}}(\mathcal{S}, T^{'})}{\mathcal{C}^{T}(t)}.
\end{equation} 
For the user-tag pair $(v,t)$, $v \in \mathcal{V}(\mathcal{G}) \setminus \mathcal{S}$, and $t \in T \setminus T^{'}$, its EMIG is defined as 
\begin{equation} \label{Eq:Node_Tag}
\delta_{(v,t)}=\frac{\sigma^{\mathcal{T}}(\mathcal{S} \cup \{v\}, T^{'} \cup \{t\})- \sigma^{\mathcal{T}}(\mathcal{S}, T^{'})}{\mathcal{C}^{S}(v) + \mathcal{C}^{T}(t)}.
\end{equation}    
\end{mydef}
Now we proceed to describe the proposed methodologies for solving the TBIM Problem.
\subsection{Methodologies Based on Effective Marginal Influence Gain Computation of User\mbox{-}Tag Pairs (EMIG-UT)} \label{Sec:UT}
In this approach, initially, the community structure of the network is detected  and then the total allocated budget for seed selection is divided among the communities based on its size (i.e., number of nodes present in it). In each community, the  shared budget is divided into two halves to be utilized to select tags and seed nodes, respectively. Next, we sort the communities based on their size in ascending order. Now, we take the smallest community first and select the most frequent tag which is less than or equal to the budget. Next, each community from smallest to the largest is processed for tag and seed node selection in the following way. Until the budget for both tag and seed node selection is exhausted, in each iteration the user\mbox{-}tag pair that causes maximum EMIG value is chosen and kept into the seed set and tag set, respectively. The extra budget for which no tag and no seed node can be selected is transferred to the largest community.  Algorithm \ref{Algo:1} describes the entire procedure.

\begin{algorithm}
\SetAlgoLined
\KwData{Social Network $\mathcal{G}(\mathcal{V}, \mathcal{E}, \mathcal{P})$, Tag Set $T$, Node Cost Function $\mathcal{C}^{S}: \mathcal{V}(\mathcal{G})\longrightarrow \mathbb{R}^{+}$, User\mbox{-}Tag Count Matrix $\mathcal{M}$, Tag Cost Function $\mathcal{C}^{T}: T \longrightarrow \mathbb{R}^{+}$, Budget $\mathcal{B}$}
\KwResult{ Seed set $\mathcal{S} \subseteq \mathcal{V} (\mathcal{G})$, and Tag set $T^{'} \subseteq T$ }
 $\mathcal{S}=\emptyset$; $T^{'}=\emptyset$ \;
 $Community=Community\_Detection(\mathcal{G})$\;
 $\mathcal{K} \longleftarrow \{\mathcal{K}_{1}, \mathcal{K}_{2}, \dots, \mathcal{K}_{\ell}\}$\;
 $\mathcal{K}_{max}=Largest\_Community(\mathcal{K})$\;
 $T_{\mathcal{K}}=Crete\_Matrix(|\mathcal{K}|,|T|, 0)$\;
 $T_{\mathcal{K}}=\text{Count the tag frequencies in each communities}$\;
 $\text{Sort row of }T_{\mathcal{K}} \text{corresponding to the smallest community }$\;
 $Create\_Vector(\mathcal{B}^{k}_{\mathcal{S}}, \ell, 0)$\;
 $Create\_Vector(\mathcal{B}^{k}_{T}, \ell, 0)$\;
\For{$i=1 \text{ to } |\mathcal{K}|$}{
 $\mathcal{B}^{k}=\frac{|\mathcal{V}_{\mathcal{K}_{i}}|}{n}. \mathcal{B}$\;
 $\mathcal{B}^{k}_{\mathcal{S}}[i]=\frac{\mathcal{B}^{k}}{2}$; $\mathcal{B}^{k}_{T}[i]=\frac{\mathcal{B}^{k}}{2}$\;
 }
 $\mathcal{K}=Sort(\mathcal{K})$\;
 $T^{'}=T^{'} \cup \{t^{'}: t^{'} \text{ is most frequent in } \mathcal{K}_{1} \text{ and } \mathcal{C}(t)\leq \mathcal{B}^{k}_{T}[i]\}$\;
 $\mathcal{B}^{k}_{T}[1]=\mathcal{B}^{k}_{T}[1]- \mathcal{C}^{T}(t^{'})$\;
 \For{$i=1 \text{ to } |\mathcal{K}|$}{
 \While{$\mathcal{B}^{k}_{\mathcal{S}}[i]>0 \text{ and }\mathcal{B}^{k}_{T}[i]>0$}{
 $(u,t)=\underset{ t^{''} \in T \setminus T^{'}, \mathcal{C}(t^{''}) \leq \mathcal{B}^{k}_{T}[i]}{\underset{v \in \mathcal{V}_{\mathcal{K}_{i}}(\mathcal{G}) \setminus \mathcal{S}, \mathcal{C}(v) \leq \mathcal{B}^{k}_{\mathcal{S}}[i];}{argmax}} \delta_{(v,t^{''})}$\;
 $\mathcal{S}=\mathcal{S} \cup \{u\}; T^{'}=T^{'} \cup \{t\}$\;
 $\mathcal{B}^{k}_{\mathcal{S}}[i]=\mathcal{B}^{k}_{\mathcal{S}}[i]-\mathcal{C}^{S}(u); \mathcal{B}^{k}_{T}[i]=\mathcal{B}^{k}_{T}[i]-\mathcal{C}^{T}(t)$\;
 }
 \footnotesize{$\mathcal{B}^{k}_{\mathcal{S}}[max]=\mathcal{B}^{k}_{\mathcal{S}}[max]+\mathcal{B}^{k}_{\mathcal{S}}[i]; \mathcal{B}^{k}_{T}[max]=\mathcal{B}^{k}_{T}[max]+\mathcal{B}^{k}_{\mathcal{S}}[i]$\;}
 }
 \caption{ Effective Marginal Influence Gain Computation of User\mbox{-}Tag Pairs (EMIG-UT)}
 \label{Algo:1}
\end{algorithm}

\par Now, we analyze Algorithm \ref{Algo:1} for its time and space requirement. Detecting communities using the Louvian Method requires $\mathcal{O}(n \log n)$ time, where $n$ denotes the number of nodes of the network \footnote{\url{https://perso.uclouvain.be/vincent.blondel/research/louvain.html}}. Computing the tag count in each community requires $\mathcal{O}(n |T|)$ time. \texttt{Community} is the array that contains the community number of the user to which they belong, i.e., \texttt{Community[i]=x} means the user $u_i$ belongs to Community $\mathcal{K}_{x}$. From this array, computing the size of each community and finding out the maximum one requires $\mathcal{O}(n)$ time. Dividing the budget among the communities for seed node and tag selection requires $\mathcal{O}(\ell)$ time. Sorting the communities requires $\mathcal{O}(\ell \log \ell)$ time. From the smallest community, choosing the highest frequency tag requires $\mathcal{O}(|T| \log |T|)$ time. The time requirement for selecting tags and seed nodes in different communities will be different. For any arbitrary community $\mathcal{K}_{x}$, let, $\mathcal{C}(\mathcal{S}^{min}_{\mathcal{K}_{x}})$ and $\mathcal{C}(T^{min}_{\mathcal{K}_{x}})$ denote the minimum seed and tag selection cost of this community, respectively. Hence, $\mathcal{C}(\mathcal{S}^{min}_{\mathcal{K}_{x}})= \underset{u \in V_{\mathcal{K}_{x}}}{min} \mathcal{C}^{S}(u)$ and $\mathcal{C}(T^{min}_{\mathcal{K}_{x}})= \underset{t \in T_{\mathcal{K}_{x}}}{min} \mathcal{C}^{T}(t)$. Here, $V_{\mathcal{K}_{x}}$ and $T_{\mathcal{K}_{x}}$ denote the nodes and relevant tags in community $\mathcal{K}_{x}$, respectively. Also, $\mathcal{B}^{\mathcal{S}}_{\mathcal{K}_{x}}$ and $\mathcal{B}^{T}_{\mathcal{K}_{x}}$ denotes the budget for selecting seed nodes and tags for the community $\mathcal{K}_{x}$, respectively. Now, it can be observed that, the number of times \texttt{while} loop (Line number $18$ to $22$) runs for the community $\mathcal{K}_{x}$ is $min\{\frac{\mathcal{B}^{\mathcal{S}}_{\mathcal{K}_{x}}}{\mathcal{C}(\mathcal{S}^{min}_{\mathcal{K}_{x}})}, \frac{\mathcal{B}^{T}_{\mathcal{K}_{x}}}{\mathcal{C}(T^{min}_{\mathcal{K}_{x}})}\}$ and it is denoted as $r_{\mathcal{K}_{x}}$. Let, $r_{max}=\underset{\mathcal{K}_{x} \in \{\mathcal{K}_{1}, \mathcal{K}_{2}, \dots, \mathcal{K}_{\ell}\}}{max} r_{\mathcal{K}_{x}}$. Hence, the number of times the marginal influence gain needs to be computed is of $\mathcal{O}(\ell r_{max}n|T|)$. Assuming the time requirement for computing the MIIA for a single node with threshold $\theta$ is of $\mathcal{O}(t_{\theta})$ \citep{wang2012scalable}. Hence, computation of $\sigma(\mathcal{S})$ requires $\mathcal{O}(nt_{\theta})$ time. Also, after updating the tag set in each iteration updating the aggregated influence probability requires $\mathcal{O}(m)$ time. Hence, execution from Line $17$ to $24$ of Algorithm \ref{Algo:1} requires $\mathcal{O}(\ell r_{max}n|T|(nt_{\theta}+m))$. Hence, the total time requirement for Algorithm \ref{Algo:1} of $\mathcal{O}(n \log n+n|T|+n +\ell+ \ell \log \ell + \ell r_{max}n|T|(nt_{\theta}+m))=\mathcal{O}(n \log n+ \ell \log \ell+\ell r_{max}n|T|(nt_{\theta}+m))$. Additional space requirement for Algorithm \ref{Algo:1} is to store the \texttt{Community} array which requires $\mathcal{O}(n)$ space, for $\mathcal{B}^{k}_{\mathcal{S}}$ and $\mathcal{B}^{k}_{T}$ require $\mathcal{O}(\ell)$, for $T_{\mathcal{K}}$ requires $\mathcal{O}(\ell |T|)$, for storing MIIA path $\mathcal{O}(n(n_{i\theta} + n_{o\theta}))$ \citep{wang2012scalable}, for aggregated influence probability $\mathcal{O}(m)$, for $\mathcal{S}$ and $T^{'}$ require $\mathcal{O}(n)$ and $\mathcal{O}(|T^{'}|)$, respectively. Formal statement is presented in Theorem \ref{Th:1}.

\begin{mytheo}\label{Th:1}
Running time and space requirement of Algorithm \ref{Algo:1} is of $\mathcal{O}(n \log n+ |T| \log |T| + \ell \log \ell+ \ell r_{max}n|T|(nt_{\theta}+m))$ and $\mathcal{O}(n(n_{i\theta} + n_{o\theta})+ \ell |T|+m)$, respectively.
\end{mytheo}
Now, we describe the changes required to Algorithm \ref{Algo:1} so that it will work for solving the TEBM Problem. First, we define the term called the `Priority of a Community'.
\begin{mydef}[Priority of a Community] \label{Def:Priority}
Let, for the given social network $\mathcal{G}(\mathcal{V}, \mathcal{E}, \mathcal{P})$, $\mathcal{K}=\{ \mathcal{K}_{1}, \mathcal{K}_{2}, \ldots, \mathcal{K}_{\ell}\}$ be the set of communities. For an $\alpha \in [0,1]$, for the community $\mathcal{K}_{i} \in \mathcal{K}$, its priority $\pi_{i}$ is defined by the Equation \ref{Eq:Priority}.
\begin{equation} \label{Eq:Priority}
    \pi_{i}= \alpha \frac{\underset{u \in \mathcal{V}_{\mathcal{K}_{i}}}{\sum} \mathcal{C}(u)}{\underset{v \in \mathcal{V}(\mathcal{G})}{\sum} \mathcal{C}(v)} + (1-\alpha)\frac{\underset{u \in \mathcal{V}_{\mathcal{K}_{i}} \cap \mathcal{D}}{\sum} b(u)}{\underset{v \in \mathcal{D}}{\sum} b(v)}
\end{equation}
\end{mydef}
The changes to Algorithm \ref{Algo:1} are as follows:
\begin{itemize}
    \item \textbf{Modification $1$:} In this case, we divide the total allocated budget among the communities based on the priorities as mentioned in Definition \ref{Def:Priority}. The reason behind this is as follows: Without loss of generality consider two communities $\mathcal{K}_{i}$ and $\mathcal{K}_{j}$. Suppose the selection costs of the nodes of Community $\mathcal{K}_{i}$ is more than that of $\mathcal{K}_{j}$. Naturally, the shared budget for the Community $\mathcal{K}_{i}$ should be more than that of $\mathcal{K}_{j}$. On the other hand, it may so happen that the total benefit that can be earned from the community $\mathcal{K}_{j}$ is more than that of $\mathcal{K}_{i}$. In this aspect, Community $\mathcal{K}_{j}$ should get more budget than the Community $\mathcal{K}_{i}$. These two are contradicting. Hence, for the TEBM Problem, we split the budget proportional to the priority of the communities. Particularly, the shared budget for the community $\mathcal{K}_{i}$ is given by the following equation:
    \begin{equation} \label{Eq:Budget_Distribution}
        \mathcal{B}^{\mathcal{K}_{i}}= \frac{\pi_{i}}{\underset{i \in [\ell]}{\sum}\pi_{i}}. \mathcal{B}
    \end{equation}
    It is easy to follow that the sum of the shared budgets of the communities will be less than equal to the total budget that has been allocated for seed set selection. 
    \item \textbf{Modification $2$:} In Line Number $19$ of Algorithm \ref{Algo:1}, the users and tags are selected based on the effective marginal influence gain. However, for solving the TEBM Problem, we select the users and tags based on the effective marginal \textbf{benefit} gain, which can be obtained by replacing $\sigma^{\mathcal{T}}(.)$ and $\beta^{\mathcal{T}}(.)$ in Equation Number \ref{Eq:Node}, \ref{Eq:Tag}, and \ref{Eq:Node_Tag}.
\end{itemize}
\par Now, we analyze the \textbf{EMIG-UT} methodology (Algorithm \ref{Algo:1} along with the suggested modifications) for solving the TEBM Problem. For computing the priorities of each of the community using Equation \ref{Eq:Priority} will require $\mathcal{O}(n)$ time. Hence, the total time requirement for computing priorities of all the communities require $\mathcal{O}(n \ell)$ time. Budget distribution among the communities using Equation \ref{Eq:Budget_Distribution} requires $\mathcal{O}(\ell)$ time. Budget distribution process of EMIG-UT method for solving the TBIM Problem (i.e., Line $10$ to $13$ of Algorithm \ref{Algo:1}) requires $\mathcal{O} (\ell)$ time. However, after incorporating the mentioned modifications this step requires $\mathcal{O}(n \ell)$ time. Still, the total time requirement remains the same. Additional space requirement after incorporating the changes is to store the priorities of the communities which requires $\mathcal{O}(\ell)$. Hence, Theorem \ref{Th:1a} holds.

\begin{mytheo}\label{Th:1a}
EMIG-UT method with the mentioned modifications can be used to solve the TEBM Problem with $\mathcal{O}(n \log n+ |T| \log |T| + \ell \log \ell+ \ell r_{max}n|T|(nt_{\theta}+m))$ time and $\mathcal{O}(n(n_{i\theta} + n_{o\theta})+ \ell |T|+m)$ space.
\end{mytheo}

\begin{algorithm}
\SetAlgoLined
\KwData{Social Network $\mathcal{G}(\mathcal{V}, \mathcal{E}, \mathcal{P})$, Tag Set $T$, Node Cost Function $\mathcal{C}^{S}: \mathcal{V}(\mathcal{G})\longrightarrow \mathbb{R}^{+}$, User\mbox{-}Tag Count Matrix $\mathcal{M}$, Tag Cost Function $\mathcal{C}^{T}: T \longrightarrow \mathbb{R}^{+}$, Budget $\mathcal{B}$}
\KwResult{ Seed set $\mathcal{S} \subseteq \mathcal{V} (\mathcal{G})$, and Tag set $T^{'} \subseteq T$ }
$\text{Execute Line Number 1 to 14 of Algorithm \ref{Algo:1}}$\;
$\text{Sort each row of the } T_{\mathcal{K}} \text{ matrix}$\;
 \For{$i=1 \text{ to } |\mathcal{K}|$}{
 \For{$j=1 \text{ to } |T|$}{
 \If{$\mathcal{C}(t_{j})\leq \mathcal{B}^{k}_{T}[i] \text{ and } t_{j} \notin T^{'}$}{
 $T^{'}=T^{'} \cup \{t_j\}$\;
 $\mathcal{B}^{k}_{T}[i]=\mathcal{B}^{k}_{T}[i] - \mathcal{C}^{T}(t_{j})$\;
 }
 }
 $\mathcal{B}_{T}^{k}[max]=\mathcal{B}_{T}^{k}[max]+\mathcal{B}^{k}_{T}[i]$\;
 }
 \For{$\text{All }(u_iu_j) \in \mathcal{E}(\mathcal{G})$}{
 $\text{Compute aggregated influence using Equation \ref{Eq:1}}$\;
 }
 \For{$i=1 \text{ to } |\mathcal{K}|$}{
 \While{$\mathcal{B}^{k}_{\mathcal{S}}[i]>0$}{
 $u=\underset{v \in \mathcal{V}_{\mathcal{K}_{i}}(\mathcal{G}) \setminus \mathcal{S}, \mathcal{C}(v) \leq \mathcal{B}^{k}_{\mathcal{S}}[i]}{argmax} \delta_{v}$\;
 $\mathcal{S}=\mathcal{S} \cup \{u\}$\;
 $\mathcal{B}^{k}_{\mathcal{S}}[i]=\mathcal{B}^{k}_{\mathcal{S}}[i]-\mathcal{C}^{S}(u)$\;
 }
 $\mathcal{B}^{k}_{\mathcal{S}}[max]=\mathcal{B}^{k}_{\mathcal{S}}[max]+\mathcal{B}^{k}_{\mathcal{S}}[i]$\;
 }
 \caption{ Effective Marginal Influence Gain Computation of User (EMIG-U)}
 \label{Algo:2}
\end{algorithm}

\subsection{Methodology Based on Effective Marginal Influence Gain Computation of  Users (EMIG-U)} \label{Sec:Users}
As observed in our experiments, computational time requirement of Algorithm \ref{Algo:1} huge, which prohibits this algorithm to be used for large scale social network datasets. To resolve this problem, Algorithm \ref{Algo:2} describes the \emph{Effective Marginal Influence Gain Computation of Users (EMIG-U)} approach, where after community detection and budget distribution, high frequency tags from the communities are chosen (Line $3$ to $11$) until budget is exhausted, and effective influence probability for each of the edges are computed. Next, from each of the communities until their respective budget is exhausted, in each iteration the node that causes maximum EMIG value are chosen as seed nodes. As described previously, time requirement for executing Line $1$ to $14$ is $\mathcal{O}(n \log n + |T| \log |T|+n |T|+n+ \ell + \ell \log \ell)=\mathcal{O}(n \log n + |T| \log |T| + n |T| + \ell \log \ell)$. Sorting each row of the matrix $T_{\mathcal{K}}$ requires $\mathcal{O}(\ell |T| \log |T|)$ time. For any arbitrary community $\mathcal{K}_{x}$, the number of times the \texttt{for} loop will run in the worst case is of $\mathcal{O}(\frac{\mathcal{B}_{\mathcal{K}_{x}}^{T}}{\mathcal{C}(T_{\mathcal{K}_{x}}^{min})})$. Let, $t_{max}=\underset{\mathcal{K}_{x} \in \{\mathcal{K}_{1}, \mathcal{K}_{2}, \dots, \mathcal{K}_{\ell}\}}{max} \frac{\mathcal{B}_{\mathcal{K}_{x}}^{T}}{\mathcal{C}(T_{\mathcal{K}_{x}}^{min})}$. Also, in every iteration, it is to be checked whether the selected tag is already in $T^{'}$ or not. Hence, the worst case running time from Line $3$ to $11$ will be of $\mathcal{O}(\ell t_{max}|T^{'}|)$. Computing aggregated influence probabilities for all the edges (Line $12$ to $14$) requires $\mathcal{O}(m |T^{'}|)$ time. For the community $\mathcal{K}_{x}$, the number of times the \texttt{while} loop in Line $16$ will run in worst case is of $\mathcal{O}(\frac{\mathcal{B}_{\mathcal{K}_{x}}^{\mathcal{S}}}{\mathcal{C}(\mathcal{S}_{\mathcal{K}_{x}}^{min})})$. Let, $s_{max}=\underset{\mathcal{K}_{x} \in \{\mathcal{K}_{1}, \mathcal{K}_{2}, \dots, \mathcal{K}_{\ell}\}}{max} \frac{\mathcal{B}_{\mathcal{K}_{x}}^{\mathcal{S}}}{\mathcal{C}(\mathcal{S}_{\mathcal{K}_{x}}^{min})}$.  Hence, the number of times the marginal influence gain will be computed is of $\mathcal{O}(\ell s_{max}n)$.
Contrary to Algorithm \ref{Algo:1}, in this case MIIA path needs to be computed only once after the tag probability aggregation is done. Hence, worst case running time from Line $15$ to $22$
is of $\mathcal{O}(\ell s_{max}n+nt_{\theta})$. The worst case running time of  Algorithm \ref{Algo:2} is of $\mathcal{O}(n \log n + |T| \log |T| + n |T| + \ell \log \ell + \ell |T| \log |T|+ \ell t_{max}|T^{'}|+ m|T^{'}|+ \ell s_{max}n+nt_{\theta})=\mathcal{O}(n \log n + n |T| + \ell \log \ell + \ell |T| \log |T|+\ell t_{max}|T^{'}|+ m|T^{'}|+ \ell s_{max}n+nt_{\theta})$. It is easy to verify that the space requirement of Algorithm \ref{Algo:2} will be same as Algorithm \ref{Algo:1}. Hence, Theorem \ref{Th:2} holds.

\begin{mytheo}\label{Th:2}
Running time and space requirement of Algorithm \ref{Algo:2} is of $\mathcal{O}(n \log n + n |T| + \ell \log \ell + \ell |T| \log |T|+\ell t_{max}|T^{'}|+ m|T^{'}|+ \ell s_{max}n+nt_{\theta})$ and $\mathcal{O}(n(n_{i\theta} + n_{o\theta})+ \ell |T|+m)$, respectively.
\end{mytheo}

Now, we describe the required changes of Algorithm \ref{Algo:2} so that it works for the TEBM Problem as well. As first $14$ lines of Algorithm \ref{Algo:1} is being executed at the beginning of Algorithm \ref{Algo:2}, so the Modification $1$ already reflects. Now, we modify Line Number $17$ of Algorithm \ref{Algo:2} as follows: instead of selecting the users based on effective marginal influence gain, we select the nodes based on the effective marginal benefit gain. As mentioned in the Section \ref{Sec:UT}, 
only the additional computational time and space requirement for solving the TEBM Problem using EMIG-U method is of $\mathcal{O}(n \ell)$ and  $\mathcal{O}(\ell)$, respectively. Hence, Theorem \ref{Th:2a} holds. 
\begin{mytheo}\label{Th:2a}
TEBM Problem can be solved using the mentioned  modifications of Algorithm \ref{Algo:2} in $\mathcal{O}(n \log n + n |T| + \ell \log \ell + \ell |T| \log |T|+\ell t_{max}|T^{'}|+ m|T^{'}|+ \ell s_{max}n+nt_{\theta})$ time and $\mathcal{O}(n(n_{i\theta} + n_{o\theta})+ \ell |T|+m)$ space.
\end{mytheo}
\subsection{Efficient Pruning Techniques}
Though, Algorithm \ref{Algo:2} has better scalability compared to Algorithm \ref{Algo:1}, still it is quite huge. To improve the scalability, here we propose efficient pruning techniques.
\subsubsection{Pruning Technique (EMIG-U-Prunn)}
 The main performance bottleneck of Algorithm \ref{Algo:2} is the excessive number of EMIG computations. Hence, it will be beneficial, if we can prune off some of the nodes, in such a way that even if we don't perform this computation for these nodes, still it does not affect much on the influence spread. We propose the following pruning strategy. Let, $\mathcal{S}^{i}$ denotes the seed set after the $i^{th}$ iteration. $\forall u \in V(\mathcal{G}) \setminus \mathcal{S}^{i}$, if the outdegree of $u$, i.e. $outdeg(u)$ will be decremented by $|\mathcal{N}^{in}(u) \cap \mathcal{S}^{i}|$, where $\mathcal{N}^{in}$ denotes the set of incoming neighbors of $u$. All the nodes in $ V (\mathcal{G}) \setminus \mathcal{S}^{i}$ are sorted based on the computed outdgree to cost ratio and top-$k$ of them are returned for the EMIG  computation. Hence, the number of times EMIG computation is happening is much reduced in this case. Instead of writing the entire algorithm once more, we only put down this pruning procedure in Algorithm \ref{Algo:Prun1}.

\begin{algorithm}[H]
  \SetAlgoLined\DontPrintSemicolon
  \SetKwFunction{algo}{algo}\SetKwFunction{proc}{proc}
  \setcounter{AlgoLine}{0}
  \SetKwProg{myproc}{Procedure}{}{}
  \myproc{EMIG-U-Pru-I($\mathcal{V} (\mathcal{G})$, $\mathcal{S}$, $\mathcal{C}$, $k$)}{
  \nl $Create\_Vector(Score, |\mathcal{V}(\mathcal{G})\setminus \mathcal{S}|, 0)$\;
  \nl \For{$\text{All } u \in \mathcal{V}(\mathcal{G}) \setminus \mathcal{S}$}{
  \nl $Score(u)= \frac{Outdegree(u)-|\mathcal{S} \cap \mathcal{N}^{in}(u)|}{\mathcal{C}(u)}$\;
  }
  \nl $\text{Sort the nodes of } \mathcal{V}(\mathcal{G})\setminus \mathcal{S} \text{ in decending order based on score value}$\;
  \nl \KwRet Top\mbox{-}k node from the sorted list. \;
  }
  \caption{Efficient Pruning Technique (EMIG-U-Prunn)}
  \label{Algo:Prun1}
\end{algorithm}

We have stated for the $i^{th}$ iteration. However, the same is performed in every iteration. Due to the space limitation, we are unable to present the entire algorithm and its analysis. However, we state the final result in Theorem \ref{Th:3}.

\begin{mytheo}\label{Th:3}
Running time and space requirement of the proposed pruning strategy is of $\mathcal{O}(n \log n + \ell |T| \log |T| + n |T| + \ell \log \ell + \ell t_{max} |T^{'}| + m |T^{'}|+ \ell s_{max}(n_{max}|\mathcal{S}|+n_{max} \log n_{max}+k) + nt_{\theta})$ and $\mathcal{O}(n(n_{i\theta} + n_{o\theta}) + \ell |T|+m)$, respectively.
\end{mytheo}
Now, for solving the TEBM Problem our pruning technique remains the same. However, this is applied on the top of the modification of Algorithm \ref{Algo:2} used for solving the TEBM Problem and discussed in Section \ref{Sec:Users}. Now, one can easily verify that the computational time and space requirement asymptotically remains the same. Hence, Theorem \ref{Th:3a} holds.
\begin{mytheo}\label{Th:3a}
 The proposed pruning strategy on the modified version of Algorithm \ref{Algo:2} can be used to solve the TEBM Problem in $\mathcal{O}(n \log n + \ell |T| \log |T| + n |T| + \ell \log \ell + \ell t_{max} |T^{'}| + m |T^{'}|+ \ell s_{max}(n_{max}|\mathcal{S}|+n_{max} \log n_{max}+k) + nt_{\theta})$ time and $\mathcal{O}(n(n_{i\theta} + n_{o\theta}) + \ell |T|+m)$ space.
\end{mytheo}
\section{Experimental Evalutions} \label{Sec:ER}
In this section, we describe the experimental evaluation of the proposed methodologies. Initially, we start with a description of the datasets.
\subsection{Datasets} 
\begin{itemize}
\item \textbf{Last.fm} \cite{Cantador:RecSys2011}: This is a popular online music platform. The dataset contains the social relation of the listener of an online music platform. 
\item \textbf{Delicious} \cite{Cantador:RecSys2011}: This is a social bookmarking web service for storing, sharing, and discovering web bookmarks.
\item \textbf{LibraryThing} \cite{cai2017spmc, zhao2015improving}: This is a book review website, where its users can make connections among themselves and share information regarding different books. 
\end{itemize}

\begin{table}[H]
  \caption{Basic Statistics of the Datasets}
  \label{tab:Dataset}
  \begin{tabular}{cccccc}
    \toprule
    Dataset Name & n & m & Density & Avg. Degree& Tags \\
    \midrule
    \textbf{Delicious} & 1288 & 11678 & 0.0070&9.06 & 11250\\
    \textbf{Last.fm} & 1839 & 25324 & 0.0075 & 13.77 & 9749\\
   \textbf{LibraryThing} & 15557 & 108987 & 0.0004 & 7.01 & 17228\\
  \bottomrule
\end{tabular}
\end{table}

Refer to Table \ref{tab:Dataset} for the basic statistics of the datasets. For all three datasets, it has been observed that the frequency of the tags decreases exponentially. Hence instead of dealing with all the tags,  we have selected $1000$ tags in each dataset using the most frequent tags per community.
\subsection{Experimental Setup}
Here, we describe the experimental setup. Initially, we start with the influence probability setting.
\subsubsection{Influence Probabilities}
\begin{itemize}

\item \textbf{Trivalency Setting}: In this case, for each edge, and for all the tags influence probabilities are randomly assigned from the set $\{0.1, 0.01, 0.001\}$.

\item \textbf{Count Probability Setting}: By this rule, for each edge $(u_iu_j)$ its influence probability vector is computed as follows. First, element-wise subtraction from $\mathcal{M}^{u_i}$ to $\mathcal{M}^{u_j}$ is performed. If there are some negative entries, they are changed to $0$. We call the obtained vector as $\mathcal{M}^{u_i-u_j}$. Next, $1$ is added with each entry of the vector $\mathcal{M}^{u_i}$. We call this vector as $\mathcal{M}^{u_i + 1}$. Now, the element-wise division of $\mathcal{M}^{u_i-u_j}$ is performed by $\mathcal{M}^{u_i + 1}$. The resultant vector is basically the influence probability vector for the edge $(u_ju_i)$. Here $1$ is added with each of the entries of $\mathcal{M}^{u_i}$ before the division just to avoid infinite values in the influence probability vector.

\item \textbf{Weighted Cascade Setting}: Let, $\mathcal{N}^{in}(u_i)$ denotes the set of incoming neighbors for the node $u_i$. In standard weighted cascade setting, $\forall u_j \in \mathcal{N}^{in}(u_i)$, the influence probability for the edges $(u_ju_i)$ is equal to $\frac{1}{deg^{in}(u_i)}$. Here, we have adopted this setting in a little different way. Let, $\mathcal{M}^{u_i}$ denotes the tag count vector of the user $u_i$ ($i^{th}$ row of the matrix $\mathcal{M}$). Now, $\forall u_j \in \mathcal{N}^{in}(u_i)$, we select the corresponding rows from $\mathcal{M}$, apply column\mbox{-}wise sum on the tag-frequency entries, and perform the element-wise division of the vector $\mathcal{M}^{u_i}$ by the summed up vector. The resultant vector is assigned as the influence probability for all the edges from $\forall u_j \in \mathcal{N}^{in}(u_i)$ to $u_i$. 
\end{itemize}
\subsubsection{Cost and Budget} We have adopted the \emph{random setting} for assigning selection costs to each user and tag as mentioned in  \cite{nguyen2013budgeted}. Selection cost for each user and tag is selected from the intervals $[50,100]$ and $[25,50]$, respectively uniformly at random. We have experimented with fixed budget values starting with $1000$, continued until $8000$, incremented each time by $1000$, i.e., $\mathcal{B}=\{1000, 2000, \ldots, 8000\}$. 
\subsubsection{Target Users:} In commercial campaigns, the item that is to be advertised must be associated with a set of features. Quite naturally, these features can be expressed with a set of tags. In this context, a user is said to be a target user if he/she is associated with at least one of these tags. We follow this technique to choose target users. Particularly, from Delicious, Last.fm, and LibraryThing datasets we choose $575$, $869$, and $13699$ many number of high frequency tags and the users corresponding to this tags are marked as target user. Details is shown in Table \ref{tab:Target_User}.

\begin{table}[H]
  \caption{Details regarding target users}
  \label{tab:Target_User}
  \begin{tabular}{c p{6.6 cm} c c}
    \toprule
    Dataset Name & Few Tags & No. of Target Users & Percentage \\
    \midrule
    \textbf{Delicious} & media awareness (id=45),  technology (id=68), internet safety (id=76), awareness (id=93),  tech (id=94), smart (id=95) & 575 & 44.64 \% \\
    \textbf{Last.fm} & rock (id=73), gothrock (id-3),  hardrock (id-72), psythedelicrock (id-75), alternative rock (id-78), glamrock (id-80), indirock (id-84),poprock(id-109) & 869 & 47 \%\\
   \textbf{LibraryThing} & love (id: 62), love story (id: 111), romance (id: 104) &  13699 & 88.05 \% \\
  \bottomrule
\end{tabular}
\end{table}

\subsubsection{Benefit Value:} We assign the benefit values to the target users from the interval $[50, 100]$ uniformly at random with an additional property described as follows. All the datasets used in our experiments contain the tag count information for every tag to every user.  
\begin{equation}
    b^{'}(v|T^{'}) = \sum_{t \in T^{'}} \sum_{u \in \mathcal{N}^{out}(v)} f(u|t) 
\end{equation}
\begin{equation}
    b(v|T^{'}) = 50 + \frac{ b^{'}(v|T^{'}) -  b^{'}_{min}(v|T^{'})}{b^{'}_{max}(v|T^{'}) -  b^{'}_{min}(v|T^{'})} * (100 -50)
\end{equation}
\begin{figure*}[t]
\centering
\begin{tabular}{ccc}
\includegraphics[scale=0.2]{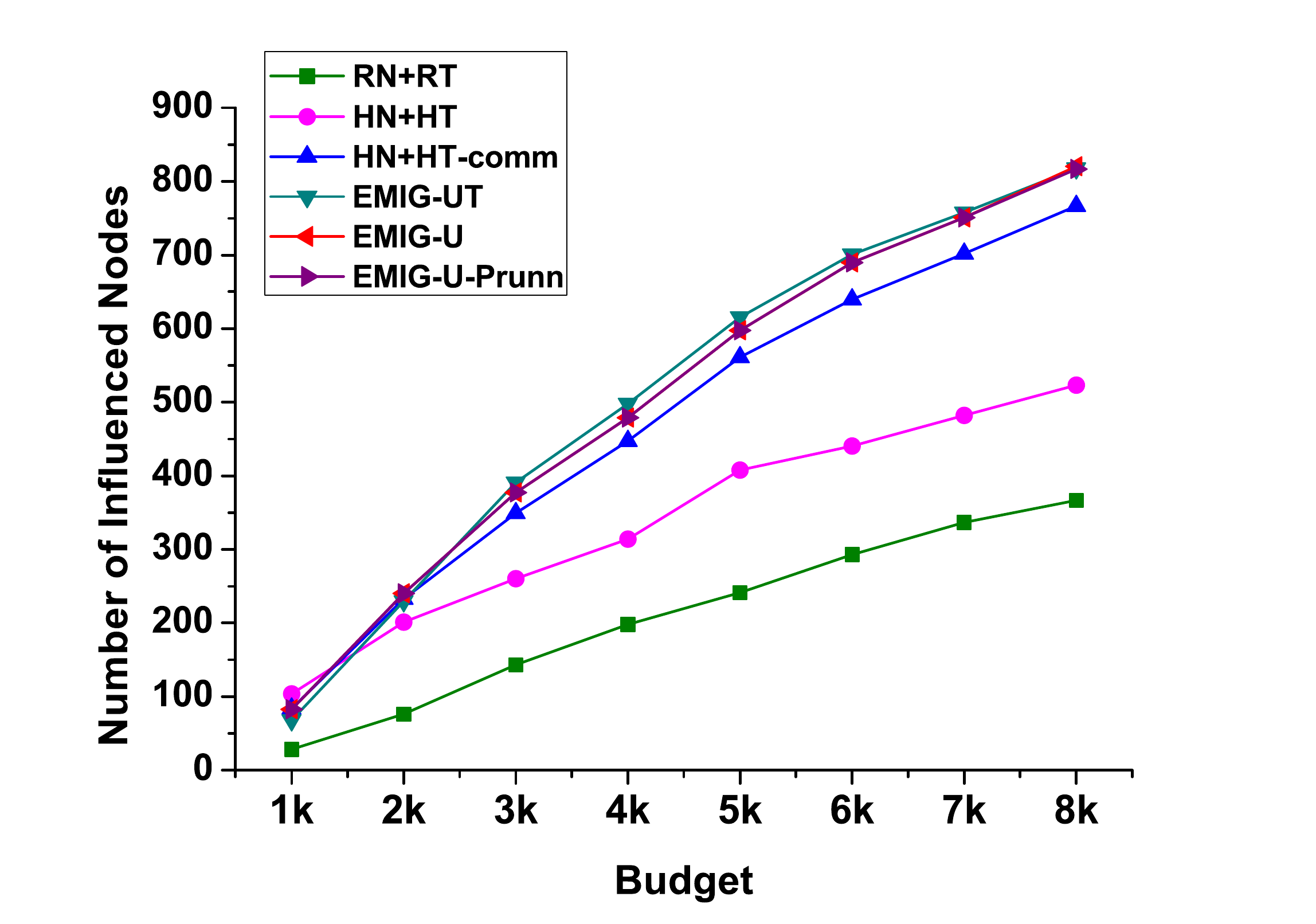} & \includegraphics[scale=0.2]{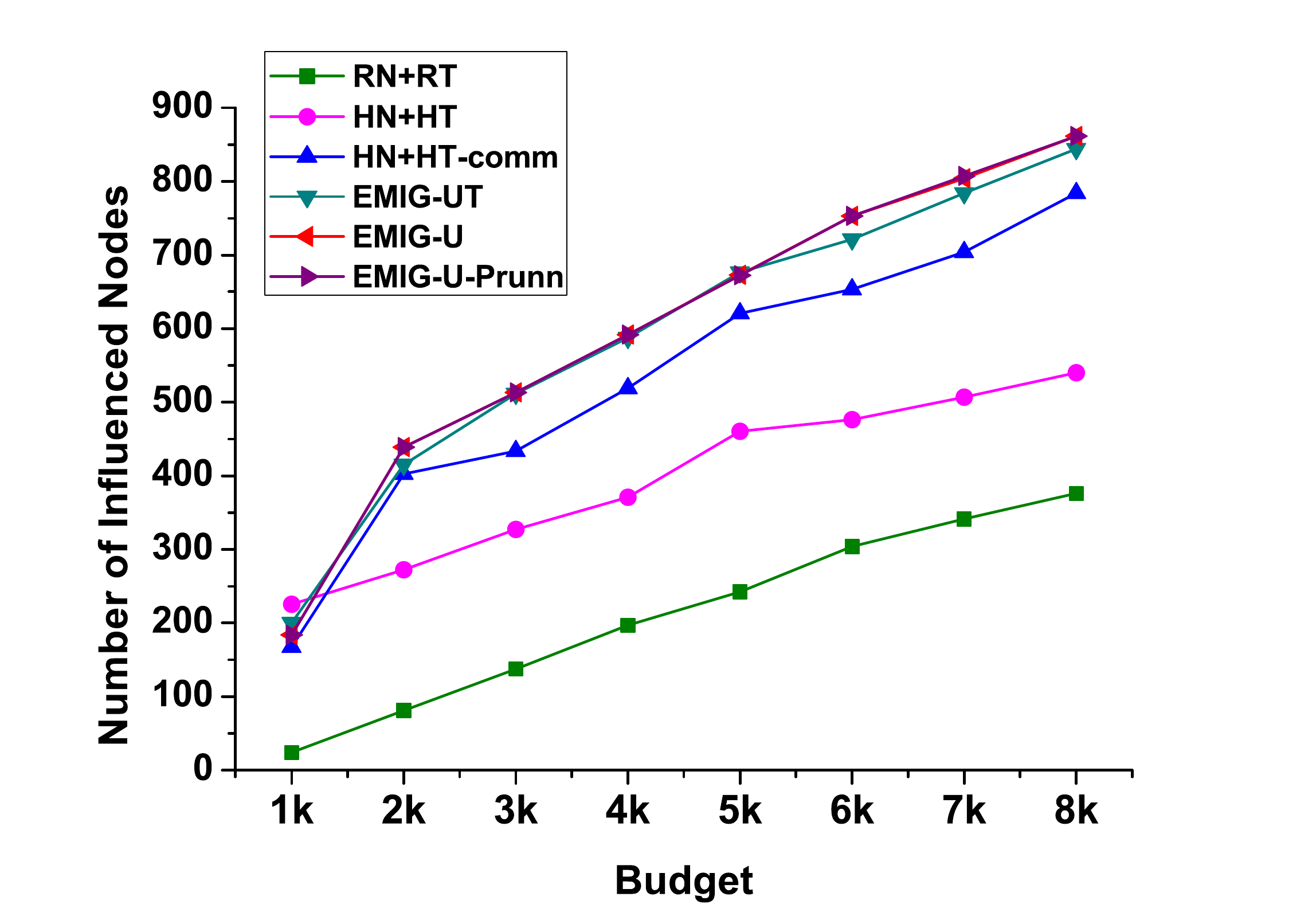} & \includegraphics[scale=0.2]{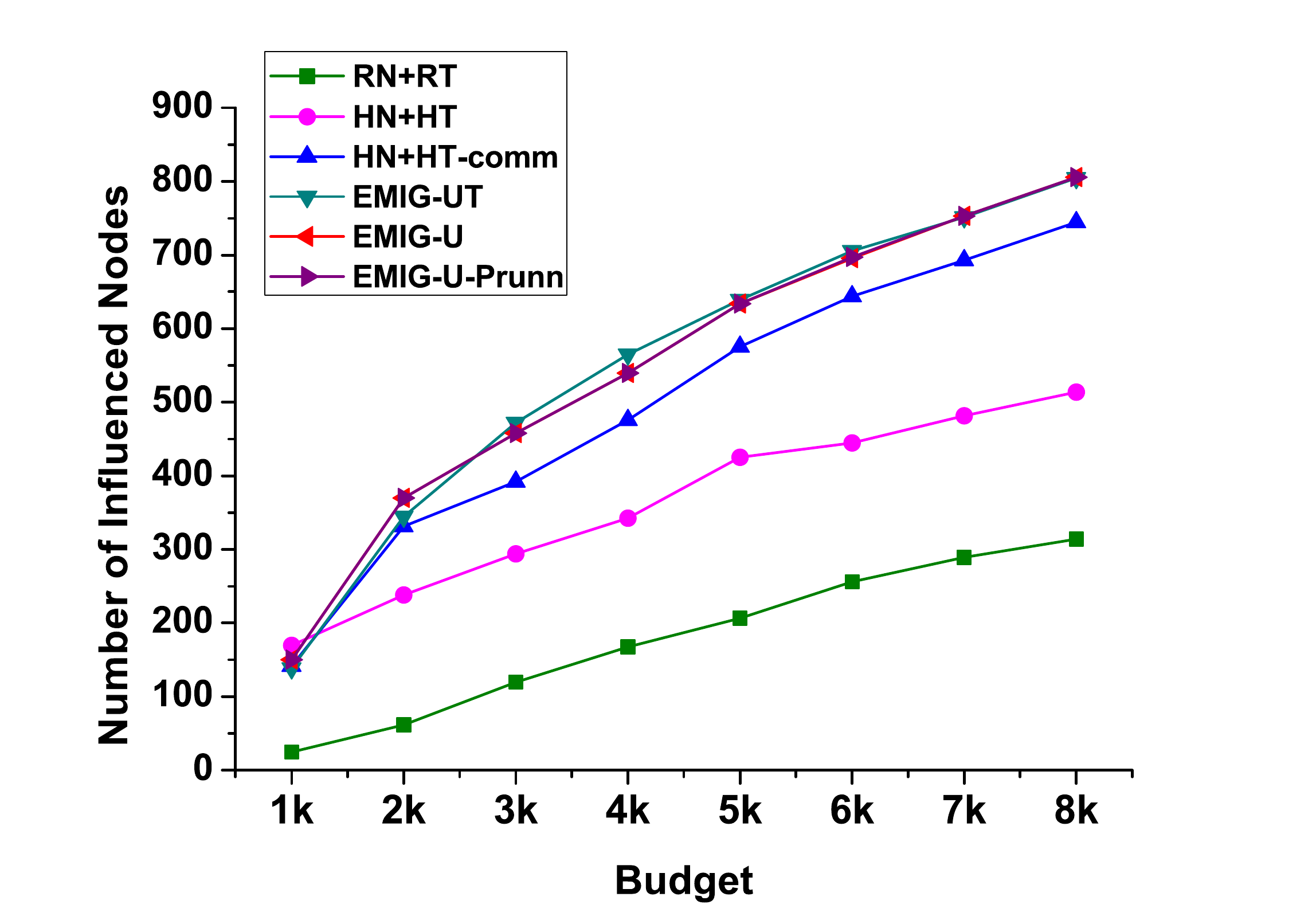}\\
(a) Delicious (Tri)  & (b) Delicious (Count) & (c) Delicious (WC) \\
\includegraphics[scale=0.2]{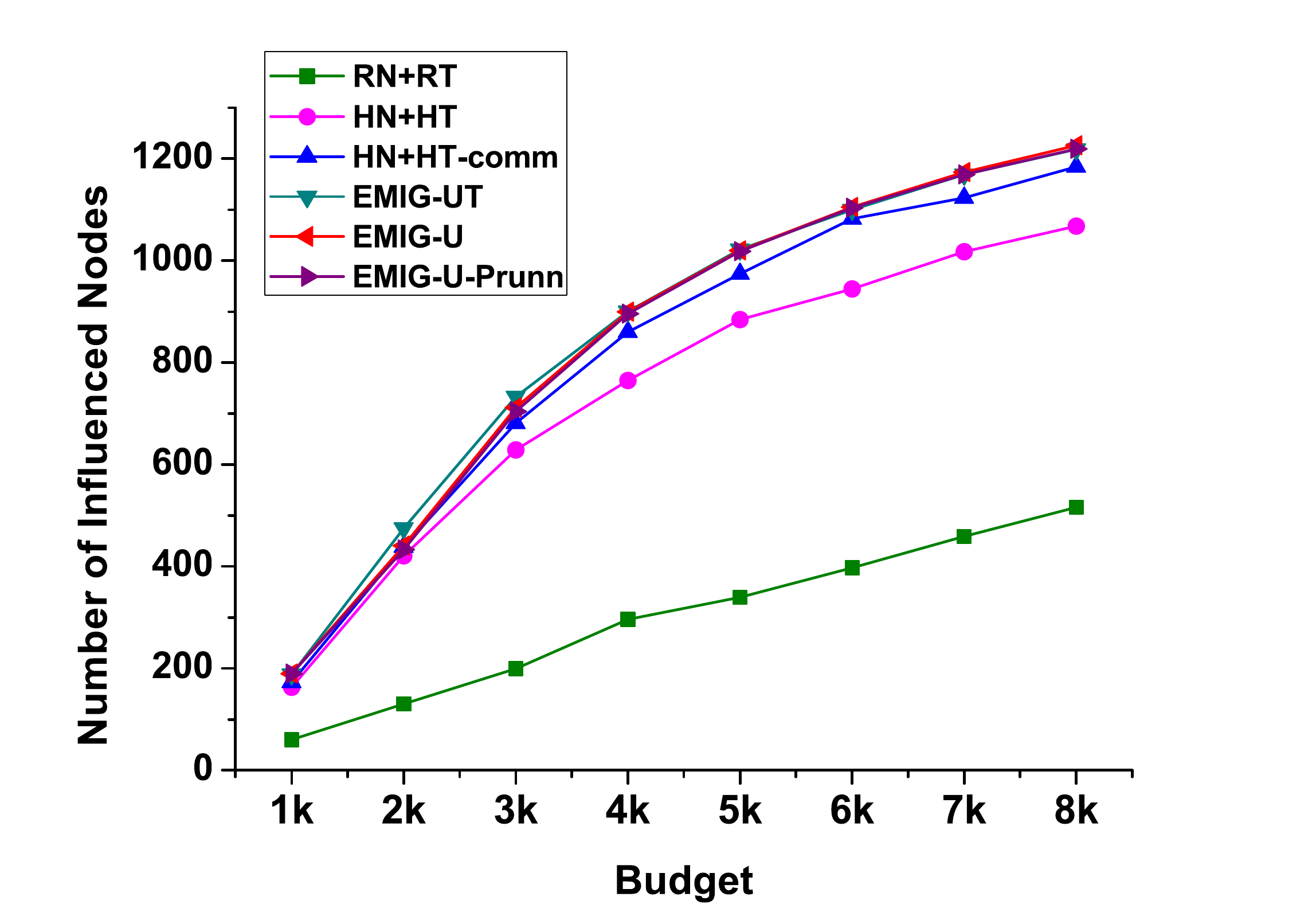} & \includegraphics[scale=0.2]{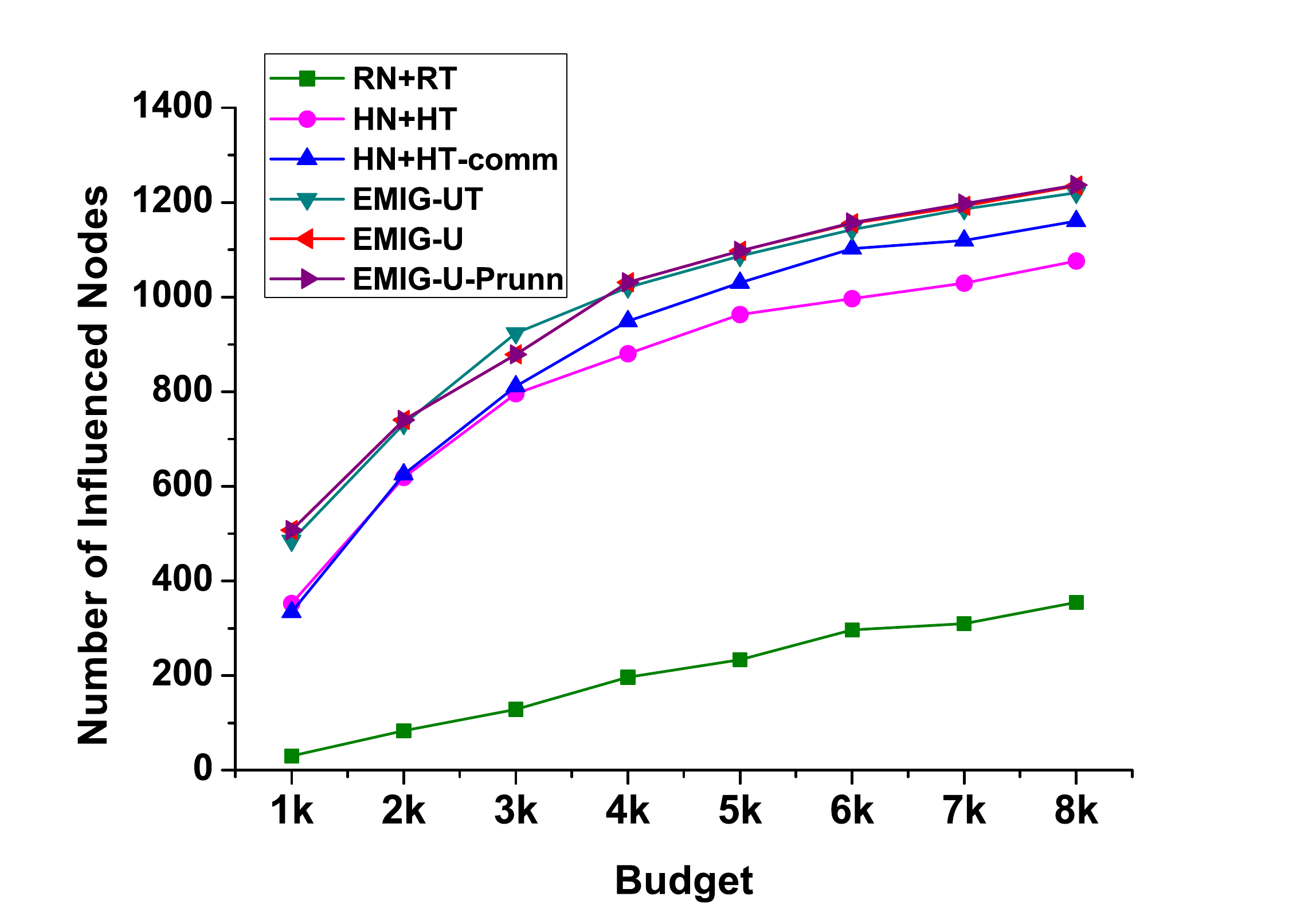} & \includegraphics[scale=0.2]{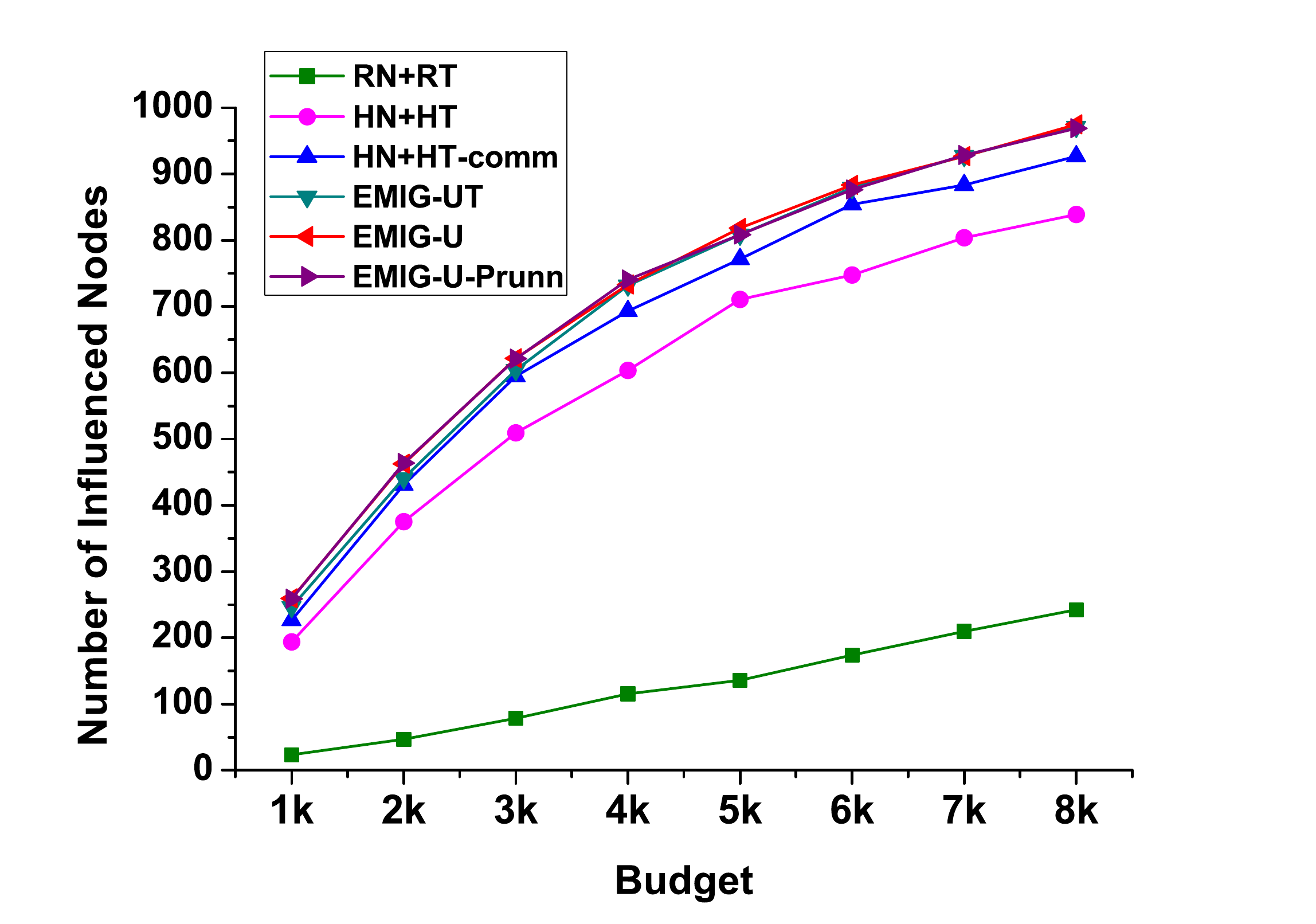}\\

(d) Last.fm (Tri) & (e) Last.fm  (Count) & (f) Last.fm  (WC) \\
\includegraphics[scale=0.2]{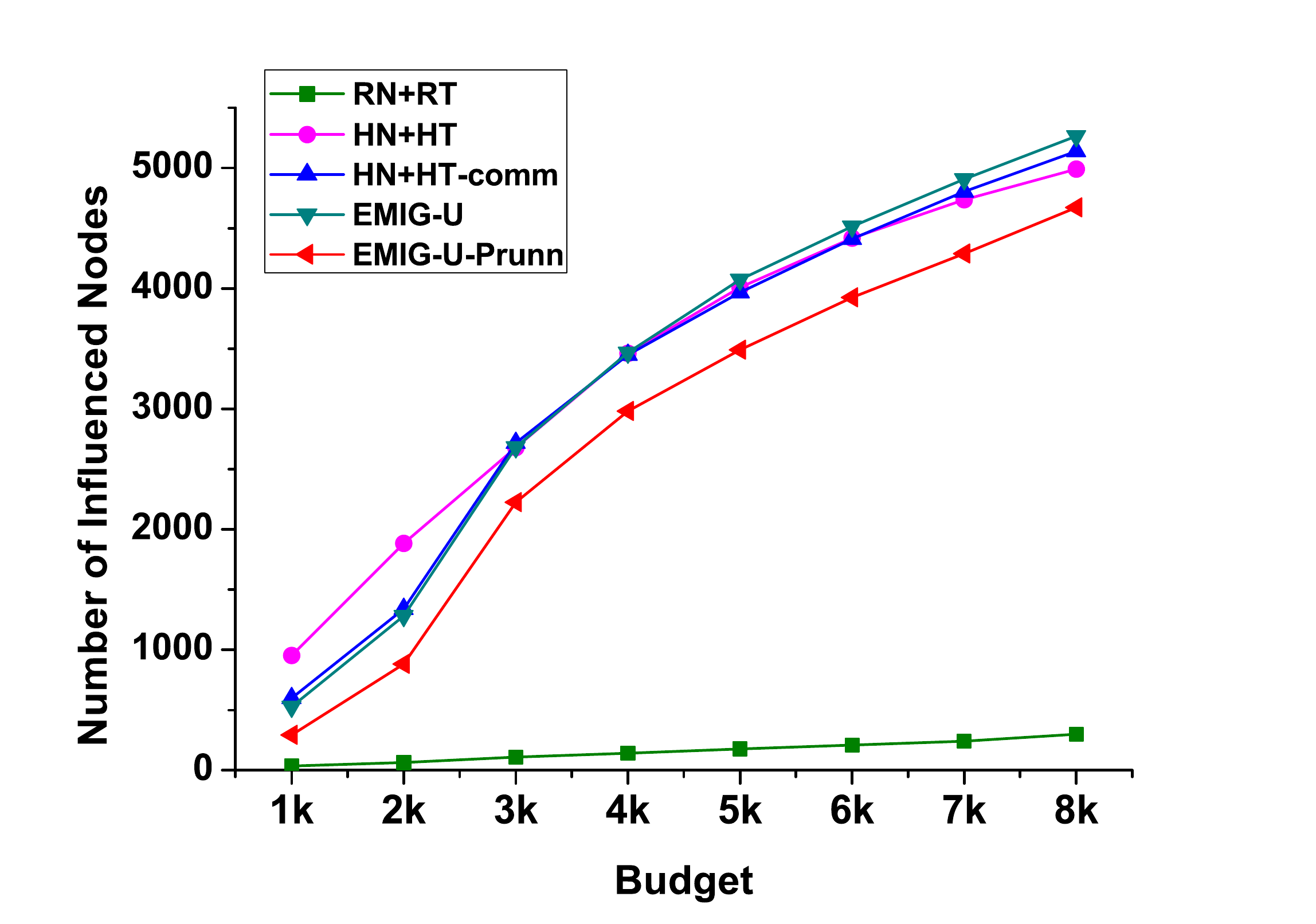} & \includegraphics[scale=0.2]{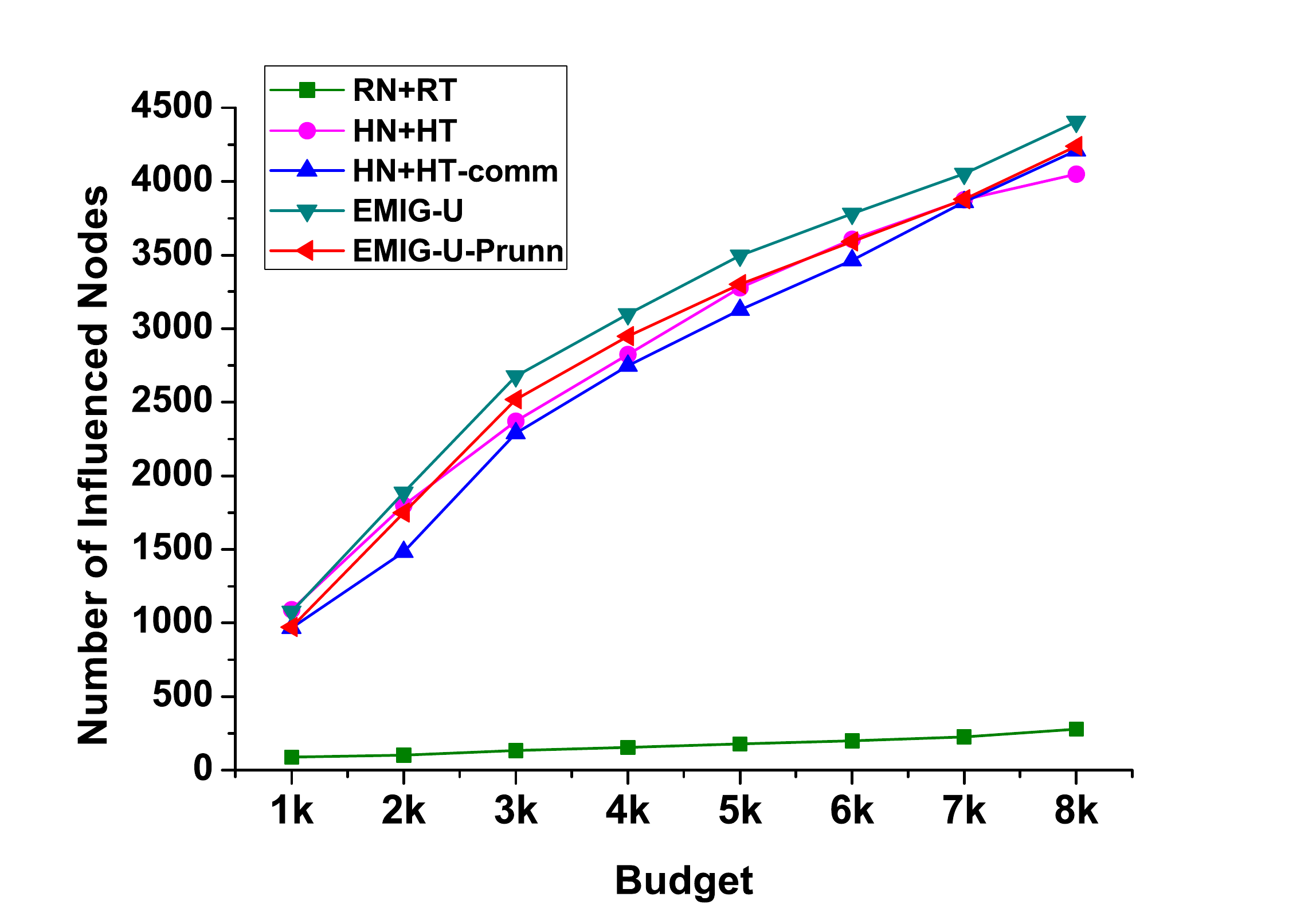} & \includegraphics[scale=0.2]{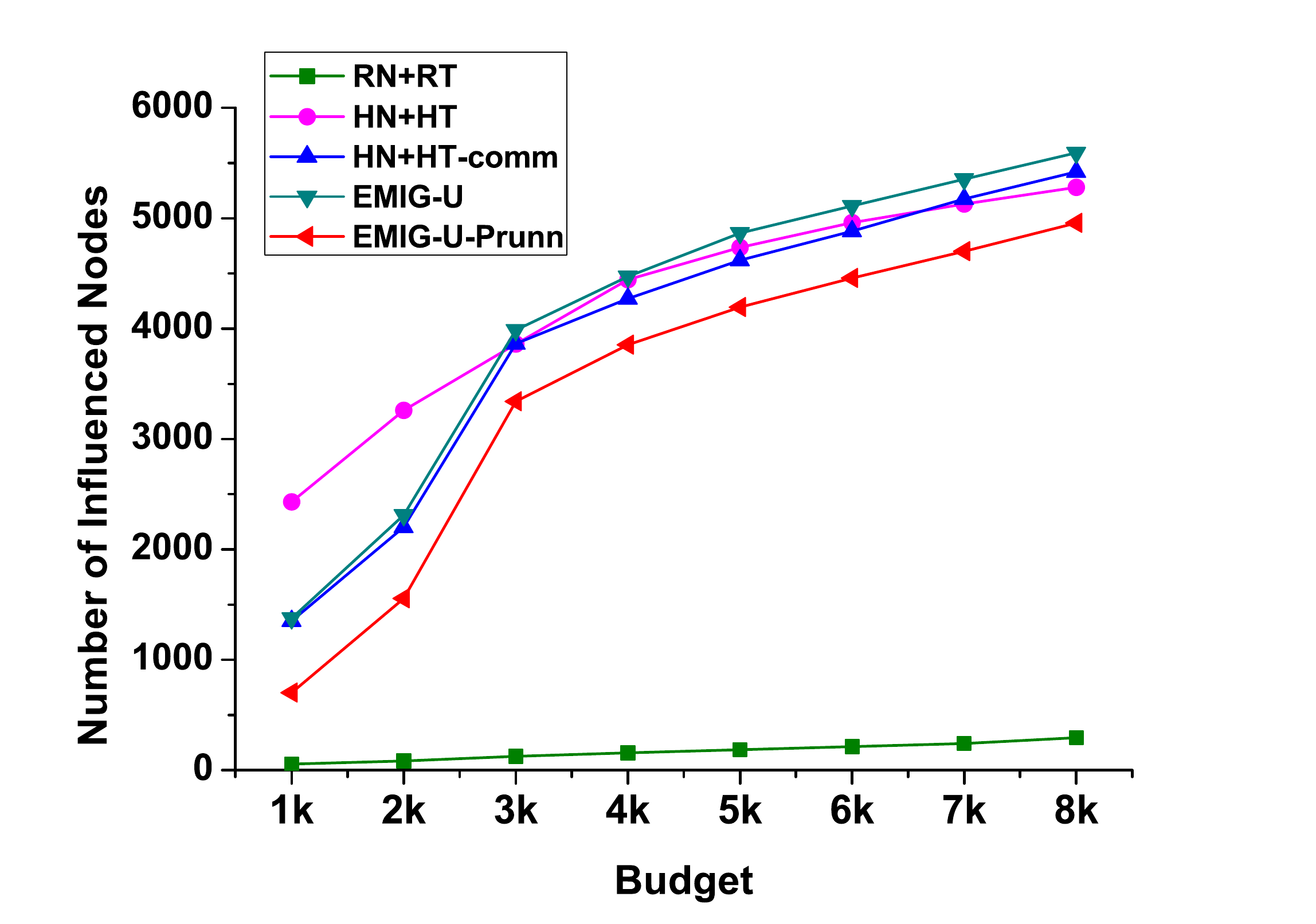}\\
(g) LibraryThing (Tri) & (h) LibraryThing (Count) & (i) LibraryThing (WC) \\
\end{tabular}
\caption{Budget Vs. Expected Influence Plot for \textbf{Delicious}, \textbf{Last.fm}, and \textbf{LibraryThing} datasets under the Trivalency, Weighted Cascade and Count Probability settings.}
\label{Fig:2}
\end{figure*}
\subsection{Algorithms Compared}
As the BIM Problem with tags has not been studied previously, there is no existing method with which we can compare the performance of the proposed methodologies. However, the following baseline methods have been used for comparison.
\begin{itemize}
\item \textbf{Random Nodes and Random Tags (RN+RT):} According to this method, the allocated budget is divided into two equal halves. One half will be spent for selecting seed nodes and the other one will be for selecting tags. Now seed nodes and tags are chosen randomly until their respective budgets are exhausted.  
\item \textbf{High Degree Nodes and High Frequency Tags (HN+HT):} According to this method, after dividing the budget into two equal halves, high degree nodes and high-frequency tags are chosen until their respective budget is exhausted. 
\item \textbf{High Degree Nodes and High Frequency Tags with Communities (HN+HT+COM):} In this method, after dividing the budget into two equal halves, first, the community structure of the network is detected. Both of these divided budgets are further divided among the communities based on the community size. Then apply \textbf{HN+HT} for each community. 
\end{itemize}
All the algorithms have been implemented with \emph{Python 3.5 + NetworkX 2.1} environment on an HPC Cluster with $5$ nodes each of them having $64$ cores and $160$ GB of memory and the implementations are available at \url{https://github.com/BITHIKA1992/BIM_with_Tag}.
\subsection{Goals of the Experimentation}
Now, we describe the goals of the experiments:
\begin{itemize}
\item to make a comparative study of the proposed as well as baseline methods of their effectiveness for solving the TBIM and TEBM Problem.
\item to make a comparative study of the proposed as well as baseline methods of their efficiency for solving the TBIM and TEBM Problem.
\item How the value of $\alpha$ has an impact on the earned benefit?
\end{itemize}
\subsection{Experimental Results with Discussions}
In this section, we describe the obtained experimental results with detailed explanations. Initially, we start by describing the impact of the seed set on influence spread.
\subsubsection{Impact on Influence Spread} \label{Sec:Inf_Spread}
Figure \ref{Fig:2} shows the Budget Vs. Expected Influence plot for all the datasets under trivalency, Count, and Weighted Cascade setting. From the figures, it has been observed that the seed set selected by proposed methodologies leads to more influence compared to the baseline methods. Now, we report the dataset\mbox{-}wise results with examples and highlight major observations. For the `Delicious' dataset, for $\mathcal{B}=8000$, under weighted cascade setting, among the baseline methods, \textbf{HN+HT+COMM} leads to the expected influence of $744$, whereas the same for \textbf{EMIG-UT}, \textbf{EMIG-U}, \textbf{EMIG-U-Prunn} methods are $804$, $805$, and $805$, respectively, which is approximately $8 \%$ more compared to \textbf{HN+HT+COMM}. The expected influence due to the seed set selected by \textbf{EMIG-U-Prunn} under Weighted Cascade, trivalency, and, Count setting are $805$, $816$, and $861$ which are $62.5\%$, $63.2\%$, and $66.85\%$ of the number of nodes of the network, respectively. Also, it is important to note, that for a given budget, the number of seed nodes selected by the proposed methodologies is always more compared to baseline methods. As an example, for $\mathcal{B}=8000$, under the trivalency setting the number of seed nodes selected by \textbf{RN+RT}, \textbf{HN+HT}, and \textbf{HN+HT+COMM} methods are $54$. However, the same for \textbf{EMIG-UT}, \textbf{EMIG-U}, and \textbf{EMIG-U-Prunn} are $59$, $62$, and $62$, respectively.
\par In the case of `Last.fm' dataset also, similar observations are made. As an example, under trivalency setting, for $\mathcal{B}=8000$, the expected influence by the proposed methodologies \textbf{EMIG-UT}, \textbf{EMIG-U}, and \textbf{EMIG-U-Prunn} is $1230$, $1226$, and $1219$, respectively. However, the same by \textbf{RN+RT}, \textbf{HN+HT}, and \textbf{HN+HT+COMM} methods are $515$, $1067$, and $1184$, respectively. In this dataset also, it has been  observed that the number of seed nodes selected by the proposed methodologies is more compared to the baseline methods. As an example, under trivalency setting, for $\mathcal{B}=8000$, the number of seed nodes selected by  \textbf{HN+HT+COMM}, and \textbf{EMIG-U-Prunn} are $51$ and $59$, respectively.
\par In this dataset, the observations are not fully consistent with the  previous two datasets. It can be observed from Figure \ref{Fig:2} ((g), (h), and (i)) that due to the pruning, the expected influence dropped significantly. As an example, in trivalency setting, for $\mathcal{B}=8000$, the expected influence by \textbf{EMIG-U} and \textbf{EMIG-U-Prunn} methods are $5265$ and $4673$, respectively. It is due to the following reason. Recall, that in the \textbf{EMIG-U-Prunn} methodology, we have only considered $200$ nodes for computing marginal gain in each iteration. As this dataset is larger than the previous two, hence there are many prospective nodes for which the marginal has not been computed. However, it is interesting to observe still the number of seed nodes selected by the proposed methodologies are more compared to baseline methods. Next, we describe the impact of the seed set on the earned benefit.

\subsubsection{Impact on Earned Benefit} \label{Sec:EB}
From the description in Section \ref{Sec:Inf_Spread} on the influence spread, it has been observed that the computational time of \textbf{EMIG-UT} is excessively huge, and hence using this algorithm for seed set selection to maximize the earned benefit in real-life situations does not make much sense. Hence, in the experiments of these sections, we do not include Algorithm \textbf{EMIG-UT}. Also, it has been observed that in different probability settings, the number of influenced nodes does not change much, hence in this section, we perform our experiments in one (in particular `count probability' setting). 

\begin{figure*}[t]
\centering
\begin{tabular}{ccc}
\includegraphics[scale=0.2]{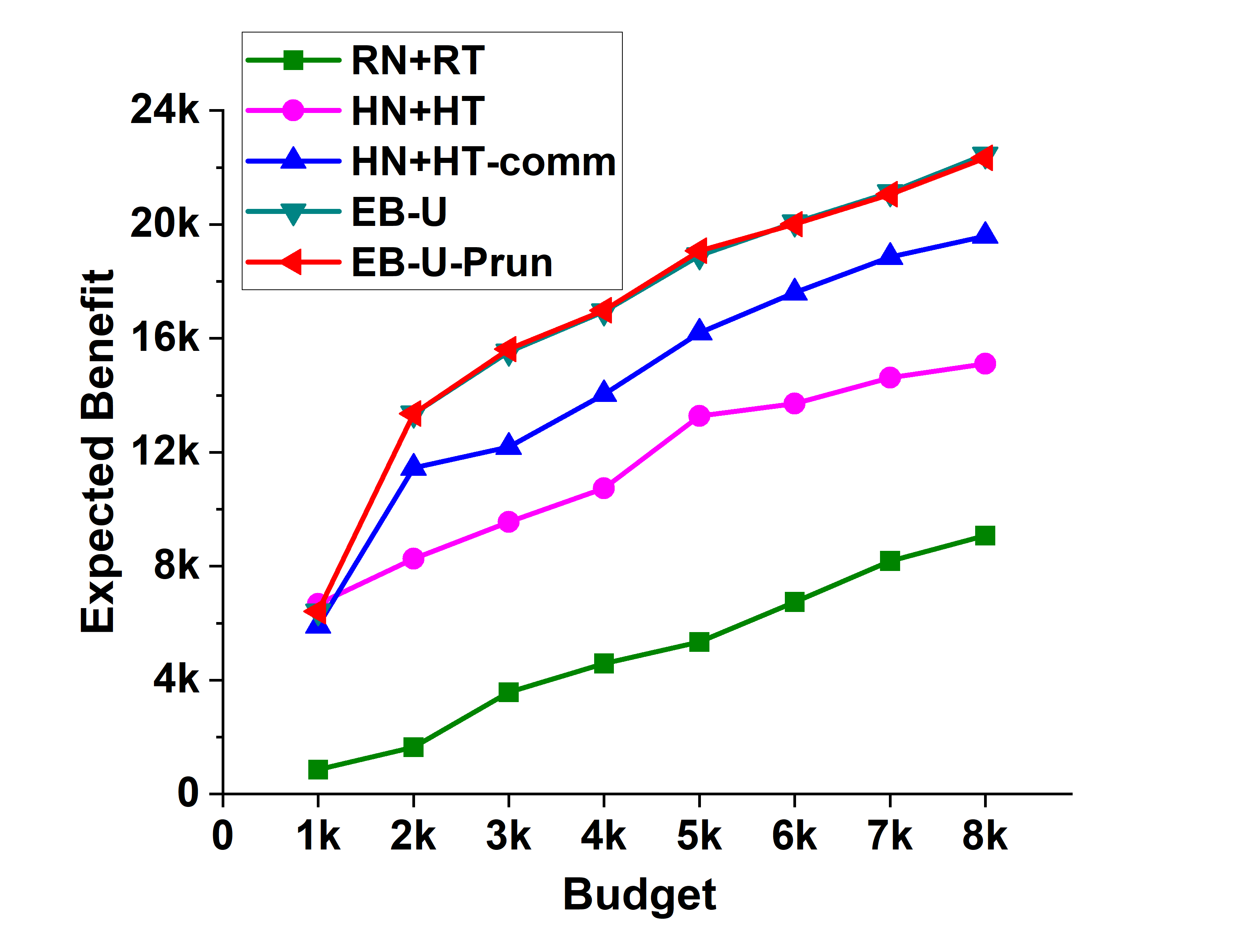} & \includegraphics[scale=0.2]{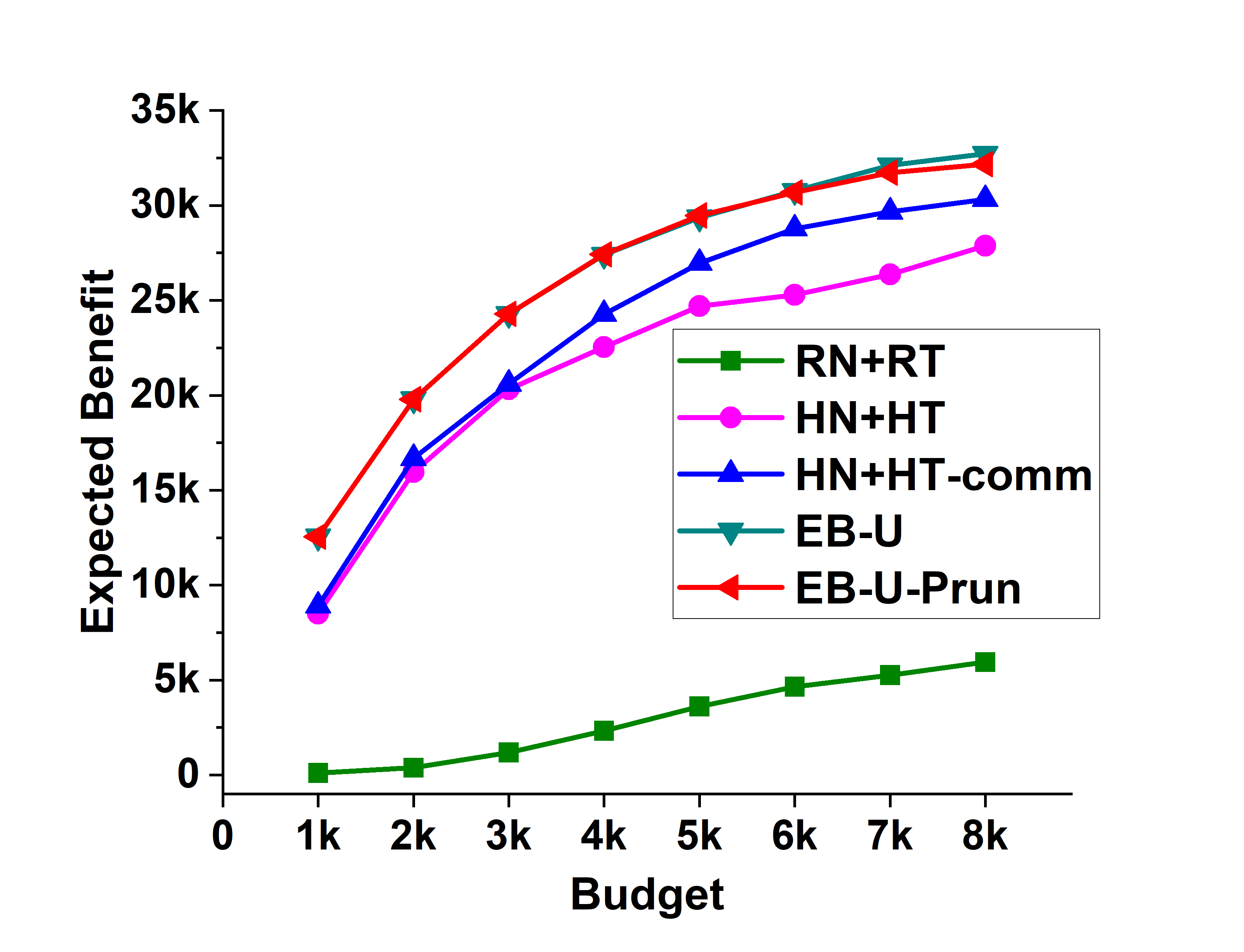} & \includegraphics[scale=0.2]{lastfm_alpha050.png}\\
(a) Delicious   & (b) Last.fm  & (c) LibraryThing \\
\end{tabular}
\caption{Budget Vs. Expected Earned Benefit Plot for \textbf{Delicious}, \textbf{Last.fm} datasets under the Count Probability settings.}
\label{Fig:3}
\end{figure*}

\par Figure \ref{Fig:3} shows the plots for the Budget vs. Expected Earned Benefit Plot for all three datasets. From the figure, it has been observed that the seed set selected by the proposed methodologies leads to more amount of expected earned benefit compared to the baseline methods. As an example for the Last.fm dataset, when $\mathcal{B}=8000$, among the baseline methods the maximum amount of earned benefit was produced  due to the seed sets selected by \textbf{HN+HT+COMM} method, and the amount is $30322$. Among the proposed methodologies \textbf{EMIG-U-Prunn} leads to the maximum amount of earned benefit which is $32171$ and this is $5.74 \%$ more compared to the \textbf{HN+HT+COMM} method. It is also important to observe that for the Last.fm dataset, for $\mathcal{B}=8000$ the earned benefit due to the proposed methodologies \textbf{EMIG-U} and \textbf{EMIG-U-Prunn} are $32714$ and $32171$, respectively which is only $1.68 \%$ less. However, the time requirement for  \textbf{EMIG-U} and \textbf{EMIG-U-Prunn} methods are $13581$ and $6452$ seconds, respectively.
\par In the case of the Delicious and the LibraryThing dataset also our observations are consistent. As an example, for the Delicious dataset when $\mathcal{B}=8000$, among the baseline methods, seed set selected by the \textbf{HN+HT+COMM} method leads to the maximum amount of earned benefit which is $19595$. In the same setting, among the proposed methodologies the earned benefit due to the seed set selected by \textbf{EMIG-U} leads to the maximum amount of earned benefit which is $22449$ and this is $0.5 \%$ more compared to the \textbf{EMIG-U-Prunn} method.
\subsubsection{Impact on Computational Time}
The computational time requirements for finding seed nodes by different methodologies are mentioned in Table \ref{Tab:Time}. It appears that among all the methods, the \textbf{RN+RT} is the fastest one. \textbf{HN+HT} requires a little more time as it incurs the cost of computing the degree of the nodes. \textbf{HN+HT+COMM} requires even more time as it incurs an  additional cost of detecting community structure. However, it is not observed for the `Delicious' dataset as it is small in size. Among the proposed methodologies, the \textbf{EMIG-UT} takes the maximum execution time among all three. This is because the number of user\mbox{-}tag combinations in each of the datasets is excessively large and marginal gain computation in each iteration incurs huge computational time. Also, in this method, after selecting the user\mbox{-}tag pair in each iteration, the effective influence probability needs to be computed for all the edges after each iteration. These are the two reasons for the excessive time requirement for \textbf{EMIG-UT}. However, \textbf{EMIG-U} is taking much less time as compared to the \textbf{EMIG-UT} due to the less number of EMIG computations. Particularly, the ratio is of order $10$ and $10^{3}$ for the Delicious and Last.fm dataset, respectively. The proposed pruning strategy reduces the computational time even further. It is observed that the ratio between the computational time of \textbf{EMIG-U} and \textbf{EMIG-U-Prunn} for the Delicious, Last.fm, and LibraryThing are approximately $1.1$, $2$, and $10$, respectively.
\par Table \ref{Tab:Time_Earned_Benefit} shows the computational time requirement for the seed set selection for the TEBM Problem. It is observed that among the baseline methods the \textbf{RN+RT} is taking the least computational time and \textbf{HN+HT+Comm} is taking maximum time. The reason is very simple. \textbf{RN+RT} method incurs the computational cost of randomly returning the nodes and tags until the budget is exhausted. On the other hand, in the case of \textbf{HN+HT+Comm} for all the node their degrees, for all the tags their counts in each of the communities needs to be computed and a community detection algorithm needs to be executed. Hence, the computational time for \textbf{HN+HT+Comm} is the highest among the baseline methods. 
\par As mentioned previously, the main bottleneck of the proposed \textbf{EMIG-U} is the excessive number of EMIG computations. However, for the \textbf{EMIG-U-Prunn} method, as the number of EMIG computations are  much less, hence this method is much faster than the \textbf{EMIG-U} method. In particular, this is much more visible for the Last.fm dataset. From Table \ref{Tab:Time_Earned_Benefit}, it can be observed that when $\mathcal{B}=8000$, the computational time requirement for selecting seed sets by \textbf{EMIG-U} and \textbf{EMIG-U-Prunn} methods are 13581.05 and  6452.45, respectively, which is approximately $2.1 \%$ more. 
\begin{table}[H]
\centering
 \caption{Computational time requirement (in Secs.) for seed set selection for the Tag\mbox{-}Based Budgeted Influence Maximization Problem}
    \label{Tab:Time}
\begin{tabular}{ | m{1.4 cm}| m{1.1 cm} | m{1.3 cm} | c | c | m{1.2 cm} | c | c |}
    \hline
    \multirow{ 2}{*}{\textbf{Dataset}} & \multirow{ 2}{*}{\textbf{Budget}} &\multicolumn{6}{|c|}{\textbf{Algorithm}}\\
    \cline{3-8} 
     &  & \textbf{EMIG-U-Prunn} & \textbf{EMIG-U} & \textbf{EMIG-UT} & \textbf{HNHTC} & \textbf{HN+HT} & \textbf{RN+RT}  \\ \hline
  \multirow{ 7}{*}{\textbf{Delicious}}  
    &1000 & 130.76  & 182.79   & $6.22 \times 10^{4}$  & 7.86 & 12.78 & 1.01  \\ \cline{2-8} 
    &2000 & 281.72  & 368.97   & $1.38 \times 10^{4}$  & 10.38 & 13.42 & 2.17  \\ \cline{2-8} 
    &3000 & 404.74  & 523.82  & $2.58 \times 10^{4}$  & 10.67  & 13.22 & 3.60   \\ \cline{2-8} 
    &4000 & 540.40  & 695.64   & $3.99 \times 10^{4}$  & 10.88 & 14.56 & 5.97  \\ \cline{2-8}
    &5000 & 704.80 & 866.16   & $4.85 \times 10^{4}$ & 11.33 & 15.26 & 8.27  \\ \cline{2-8}
    &6000 & 798.86  & 909.35  & $5.13 \times 10^{4}$  & 11.31  & 13.81 & 11.93\\ \cline{2-8}
    &7000 & 926.89  & 1068.15  & $6.92\times 10^{4}$  & 11.36  & 13.27 & 13.46 \\ \cline{2-8}
    &8000 & 1085.05 & 1259.23 & $6.82\times 10^{4}$  & 11.49  & 14.40 & 11.94  \\ 
    
    \hline
    
      \multirow{ 7}{*}{\textbf{Last.fm}}   &   1000 & 305.56  & 712.28 & $2.23 \times 10^{5}$ & 31.37 & 21.05 & 1.85 \\ \cline{2-8} 
    &2000 & 919.89  & 1986.23 & $8.14 \times 10^{5}$ & 40.42 & 28.34 & 6.83  \\ \cline{2-8} 
    &3000 & 1496.92  & 2991.10 & $1.38 \times 10^{6}$ & 50.79 & 31.73 & 8.80 \\ \cline{2-8} 
    &4000 & 2099.94 & 4160.16 & $1.92 \times 10^{6}$ & 57.83 & 38.54 & 11.29  \\ \cline{2-8} 
    &5000 & 2540.25 & 5028.90  & $2.16 \times 10^{6}$ & 61.42 & 43.80 & 13.49  \\ \cline{2-8} 
    &6000 & 3101.83 & 6381.85  & $2.42 \times 10^{6}$ & 67.92 & 47.37 & 16.74  \\  \cline{2-8} 
    &7000 & 3613.09 & 7248.25  & $2.74 \times 10^{6}$ & 69.97 & 50.62 & 20.89  \\  \cline{2-8}
    &8000 & 4032.52 & 7954.09 & $2.91 \times 10^{6}$  & 70.32 & 50.99 & 22.40  \\
    \hline
      \multirow{ 7}{*}{ \textbf{Library}}  &    1000 & 1792.85 & $2.52 \times 10^{4}$ &  N.A. & 134.20 & 89.95 & 174.40 \\ \cline{2-8} 
   & 2000 & 4870.87 & $6.35 \times 10^{4}$ & N.A. & 222.54 & 129.36 & 234.08 \\ \cline{2-8} 
   &3000 & 9328.23 & $1.17 \times 10^{5}$ & N.A. & 241.19 & 170.45 & 265.36 \\ \cline{2-8} 
   & 4000 & 13193.17 & $1.65 \times 10^{5}$ & N.A. & 288.62 & 185.27 & 308.76  \\ \cline{2-8} 
   & 5000 & 17242.24 & $2.03 \times 10^{5}$ & N.A. & 312.22 & 218.72 & 342.88 \\ \cline{2-8} 
   & 6000 & 21705.09 & $2.48 \times 10^{5}$ & N.A. & 331.24 & 232.68 & 344.22 \\ \cline{2-8} 
   & 7000 & 25269.24 & $2.84 \times 10^{5}$ & N.A. & 365.39 & 264.74 & 379.19 \\ \cline{2-8}
   & 8000 & 29544.79 & $3.26 \times 10^{5}$ & N.A. & 390.14 & 264.54 & 386.04 \\
    \hline
    \end{tabular}

\end{table}

\begin{table}[H]
\centering
 \caption{Computational time requirement (in Secs.) for seed set selection for the Tag\mbox{-}Based Earned Benefit Maximization Problem}
    \label{Tab:Time_Earned_Benefit}
\begin{tabular}{ | m{1.4 cm}| m{1.1 cm} | m{1.3 cm} | c | m{1.4 cm} | c | c |}
    \hline
    \multirow{ 2}{*}{\textbf{Dataset}} & \multirow{ 2}{*}{\textbf{Budget}} &\multicolumn{5}{|c|}{\textbf{Algorithm}}\\
    \cline{3-7} 
     &  & \textbf{EMIG-U-Prunn} & \textbf{EMIG-U} & \textbf{HNHTC} & \textbf{HN+HT} & \textbf{RN+RT}  \\ \hline
  \multirow{ 7}{*}{\textbf{Delicious}}  
    &1000 & 174.10  &  243.60  &   7.40  & 7.76 &  1.64 \\ \cline{2-7} 
    &2000 & 447.12  & 622.62 &  10.18  & 9.809 & 3.31  \\ \cline{2-7} 
    &3000 & 624.97 &  873.46 &  10.44  & 10.69 & 6.35 \\ \cline{2-7} 
    &4000 & 831.82 &  1442.17&  11.26  & 10.71 &  7.40 \\ \cline{2-7}
    &5000 & 1020.15 & 1275.10 &  11.29 & 11.19 & 8.19 \\ \cline{2-7}
    &6000 & 1335.74 &  1571.07 &   11.52  &  11.27& 8.39\\ \cline{2-7}
    &7000 & 1505.23 & 1788.42  & 11.64  & 11.66 &  9.34\\ \cline{2-7}
    &8000 & 1630.12 & 1952.08 & 11.87 & 11.69 & 9.65 \\ 
    
    \hline
    
      \multirow{ 7}{*}{\textbf{Last.fm}}   &   1000 & 456.09 & 1199.66 & 24.65 & 23.74 & 1.37 \\ \cline{2-7} 
    &2000 & 1227.08 & 2854.29 & 33.30 & 33.06 &  3.19 \\ \cline{2-7} 
    &3000 &  2029.02 & 4661.47 & 40.61 & 37.31 & 4.01 \\ \cline{2-7} 
    &4000 & 2920.05 & 6311.78 & 49.87 & 44.96 & 4.68  \\ \cline{2-7} 
    &5000 & 3806.55 & 8099.59  & 54.22 & 50.03 &  10.43 \\ \cline{2-7} 
    &6000 & 4527.61 & 9695.24 & 56.99  & 53.24 & 11.95  \\  \cline{2-7} 
    &7000 & 5413.39 & 11620.07 &  58.01  & 59.76 &  15.20 \\  \cline{2-7}
    &8000 & 6452.45 & 13581.05 &  61.72 & 58.80 &  15.70 \\
    \hline
    \end{tabular}

\end{table}
\subsubsection{Impact of $\alpha$}
\begin{figure*}[t]
\centering
\begin{tabular}{cc}
\includegraphics[scale=0.2]{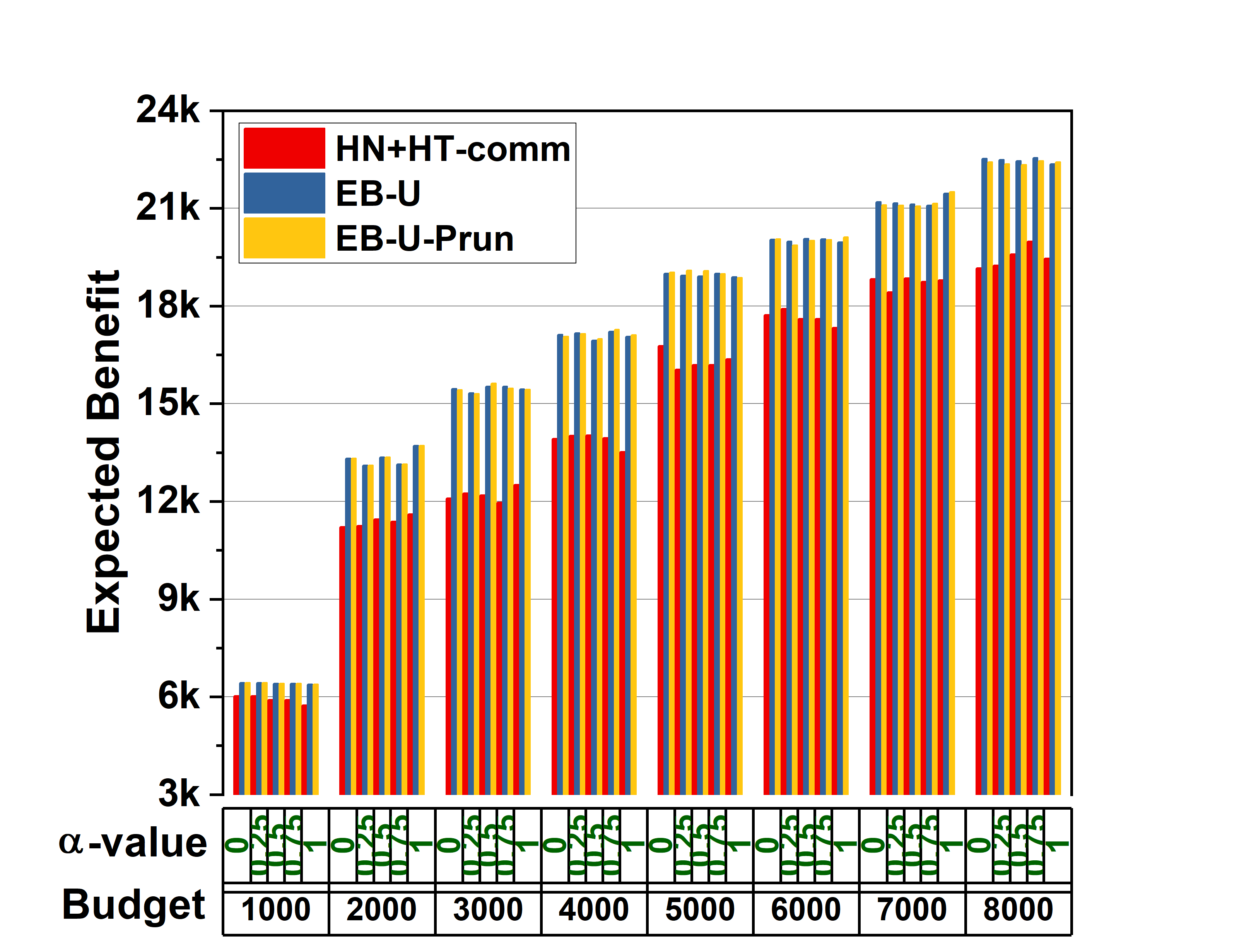} & \includegraphics[scale=0.2]{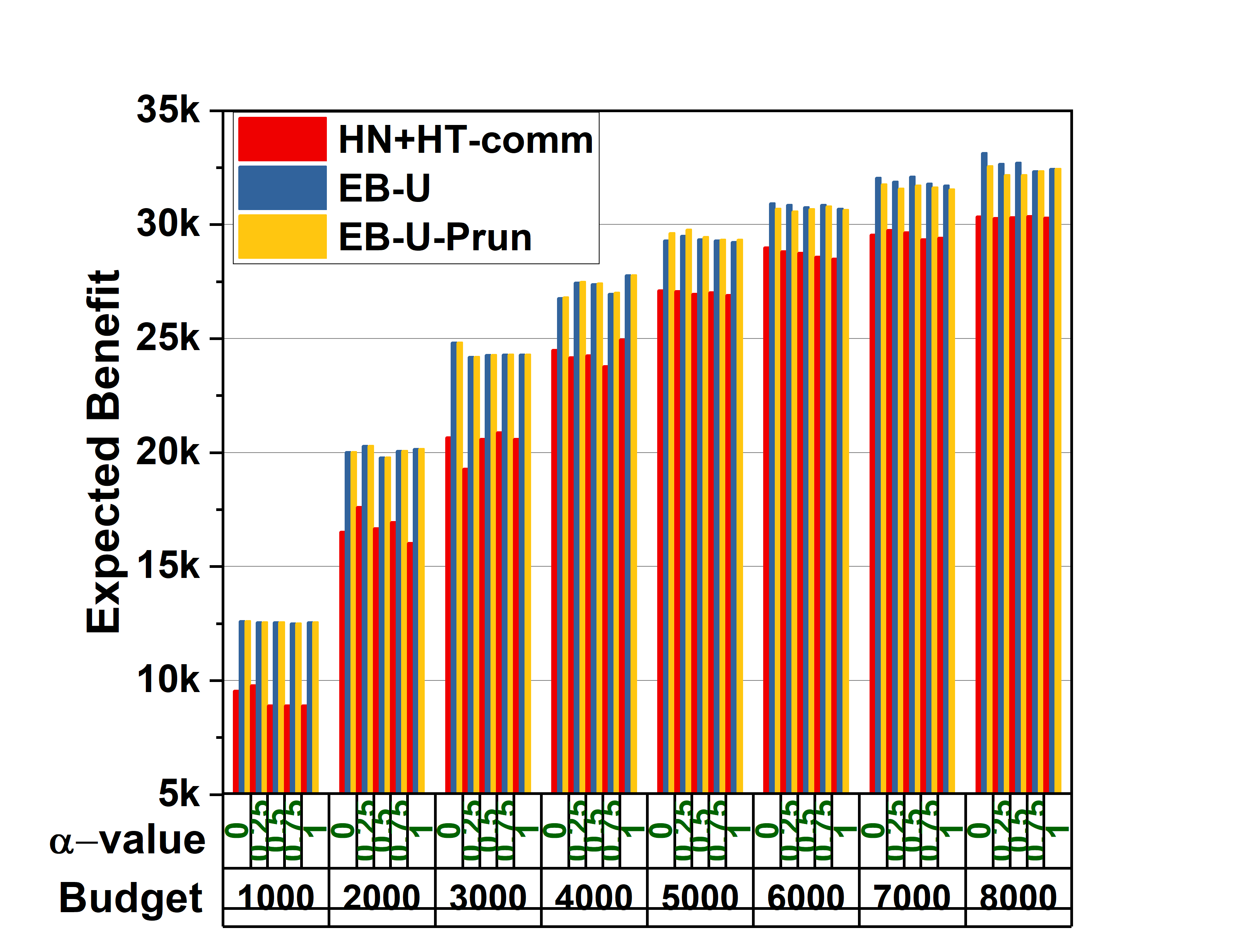} \\
(a) Delicious   & (b) Last.fm   \\
\end{tabular}
\caption{Impact of $\alpha$ on Earned Benefit on Delicious and Last.fm Dataset}
\label{Fig:4}
\end{figure*}

In this section, we describe the impact of $\alpha$ on the earned benefit. From Equation 12, it is easy to understand that $\alpha$ tries to balance between the cost of the nodes in a community with their corresponding benefits, to prioritize the budget allocation for a community. If $\alpha = 1$, only the cost of the nodes contributes to the budget allocation, whereas in the case of $\alpha = 0$, the benefits of the users are considered. For both delicious and last.fm datasets, the results of expected earned benefit for different $\alpha ( e.g., 0.0, 0.25, 0.5, 0.75, 1.0)$ values are shown in Figure \ref{Fig:4}. The cost of the users is chosen uniformly at random within the range of 50 to 100. The benefit is also selected within the same range of cost, and it returns zero if neither the user nor his neighbor selects at least one targeted tag. For all the methods which use community-based budget allocation, the change in expected benefits are visible for smaller values of budget and it is maximum for $\alpha = zero$. Compared to HN+HT-Comm, the expected benefit-based algorithms are more uniform for different values of $\alpha$, due to their uniform choice of cost and benefits range. One interesting point to notice here is that in both the datasets, the earned benefit gaps within HN+HT-Comm and expected-benefit based algorithms are high for large value of $\alpha$, i.e., when the budget is almost dependent on the cost of users, the high degree node does not incur the maximum expected benefit. 

\section{Conclusion} \label{Sec:CFD}
In this paper, we have introduced the Tag\mbox{-}Based Budgeted Influence Maximization Problem and the Tag\mbox{-}Based Earned Benefit Maximization Problem. For both these problems, considering the ground truth that different tags have a different level of popularity in different communities, three different methodologies have been proposed. All the methodologies have been analyzed for their time and space requirements. From the experimental results, it has been observed that the seed set selected by the proposed methodologies leads to better influence spread compared to the baseline methods. Our future study on this problem will remain concentrated on developing efficient pruning  techniques, which reduces the computational time without affecting the influence spread much. A number of avenues are open to extend this work. In this study, we have not considered the presence of negative information spreader. Our study can be enhanced by considering this situation. Also, efficient pruning techniques can be proposed such that the proposed methodologies will be much more efficient.

\bibliography{sample}

\begin{thebibliography}{52}
\expandafter\ifx\csname natexlab\endcsname\relax\def\natexlab#1{#1}\fi
\providecommand{\url}[1]{\texttt{#1}}
\providecommand{\href}[2]{#2}
\providecommand{\path}[1]{#1}
\providecommand{\DOIprefix}{doi:}
\providecommand{\ArXivprefix}{arXiv:}
\providecommand{\URLprefix}{URL: }
\providecommand{\Pubmedprefix}{pmid:}
\providecommand{\doi}[1]{\href{http://dx.doi.org/#1}{\path{#1}}}
\providecommand{\Pubmed}[1]{\href{pmid:#1}{\path{#1}}}
\providecommand{\bibinfo}[2]{#2}
\ifx\xfnm\relax \def\xfnm[#1]{\unskip,\space#1}\fi
\bibitem[{Banerjee et~al.(2019a)Banerjee, Jenamani and
  Pratihar}]{banerjee2019combim}
\bibinfo{author}{Banerjee, S.}, \bibinfo{author}{Jenamani, M.},
  \bibinfo{author}{Pratihar, D.K.}, \bibinfo{year}{2019}a.
\newblock \bibinfo{title}{Combim: A community-based solution approach for the
  budgeted influence maximization problem}.
\newblock \bibinfo{journal}{Expert Systems with Applications} .
\bibitem[{Banerjee et~al.(2019b)Banerjee, Jenamani and
  Pratihar}]{banerjee2019maximizing}
\bibinfo{author}{Banerjee, S.}, \bibinfo{author}{Jenamani, M.},
  \bibinfo{author}{Pratihar, D.K.}, \bibinfo{year}{2019}b.
\newblock \bibinfo{title}{Maximizing the earned benefit in an incentivized
  social networking environment: a community-based approach}.
\newblock \bibinfo{journal}{Journal of Ambient Intelligence and Humanized
  Computing} , \bibinfo{pages}{1--17}.
\bibitem[{Banerjee et~al.(2019c)Banerjee, Jenamani and
  Pratihar}]{banerjee2019maximizingw}
\bibinfo{author}{Banerjee, S.}, \bibinfo{author}{Jenamani, M.},
  \bibinfo{author}{Pratihar, D.K.}, \bibinfo{year}{2019}c.
\newblock \bibinfo{title}{Maximizing the earned benefit in an incentivized
  social networking environment: An integer programming-based approach}, in:
  \bibinfo{booktitle}{Proceedings of the ACM India Joint International
  Conference on Data Science and Management of Data}, pp.
  \bibinfo{pages}{322--325}.
\bibitem[{Banerjee et~al.(2020a)Banerjee, Jenamani and
  Pratihar}]{banerjee2020earned}
\bibinfo{author}{Banerjee, S.}, \bibinfo{author}{Jenamani, M.},
  \bibinfo{author}{Pratihar, D.K.}, \bibinfo{year}{2020}a.
\newblock \bibinfo{title}{Earned benefit maximization in social networks under
  budget constraint}.
\newblock \bibinfo{journal}{Expert Systems with Applications} ,
  \bibinfo{pages}{114346}.
\bibitem[{Banerjee et~al.(2020b)Banerjee, Jenamani and
  Pratihar}]{banerjee2020survey}
\bibinfo{author}{Banerjee, S.}, \bibinfo{author}{Jenamani, M.},
  \bibinfo{author}{Pratihar, D.K.}, \bibinfo{year}{2020}b.
\newblock \bibinfo{title}{A survey on influence maximization in a social
  network}.
\newblock \bibinfo{journal}{Knowledge and Information Systems} ,
  \bibinfo{pages}{1--39}.
\bibitem[{Banerjee et~al.(2020c)Banerjee, Pal and
  Jenamani}]{banerjee2020budgeted}
\bibinfo{author}{Banerjee, S.}, \bibinfo{author}{Pal, B.},
  \bibinfo{author}{Jenamani, M.}, \bibinfo{year}{2020}c.
\newblock \bibinfo{title}{Budgeted influence maximization with tags in social
  networks}, in: \bibinfo{booktitle}{International Conference on Web
  Information Systems Engineering}, \bibinfo{organization}{Springer}. pp.
  \bibinfo{pages}{141--152}.
\bibitem[{Bozorgi et~al.(2016)Bozorgi, Haghighi, Zahedi and
  Rezvani}]{bozorgi2016incim}
\bibinfo{author}{Bozorgi, A.}, \bibinfo{author}{Haghighi, H.},
  \bibinfo{author}{Zahedi, M.S.}, \bibinfo{author}{Rezvani, M.},
  \bibinfo{year}{2016}.
\newblock \bibinfo{title}{Incim: A community-based algorithm for influence
  maximization problem under the linear threshold model}.
\newblock \bibinfo{journal}{Information Processing \& Management}
  \bibinfo{volume}{52}, \bibinfo{pages}{1188--1199}.
\bibitem[{Bozorgi et~al.(2017)Bozorgi, Samet, Kwisthout and
  Wareham}]{bozorgi2017community}
\bibinfo{author}{Bozorgi, A.}, \bibinfo{author}{Samet, S.},
  \bibinfo{author}{Kwisthout, J.}, \bibinfo{author}{Wareham, T.},
  \bibinfo{year}{2017}.
\newblock \bibinfo{title}{Community-based influence maximization in social
  networks under a competitive linear threshold model}.
\newblock \bibinfo{journal}{Knowledge-Based Systems} \bibinfo{volume}{134},
  \bibinfo{pages}{149--158}.
\bibitem[{Buchbinder et~al.(2015)Buchbinder, Feldman, Seffi and
  Schwartz}]{buchbinder2015tight}
\bibinfo{author}{Buchbinder, N.}, \bibinfo{author}{Feldman, M.},
  \bibinfo{author}{Seffi, J.}, \bibinfo{author}{Schwartz, R.},
  \bibinfo{year}{2015}.
\newblock \bibinfo{title}{A tight linear time (1/2)-approximation for
  unconstrained submodular maximization}.
\newblock \bibinfo{journal}{SIAM Journal on Computing} \bibinfo{volume}{44},
  \bibinfo{pages}{1384--1402}.
\bibitem[{Bucur and Iacca(2016)}]{bucur2016influence}
\bibinfo{author}{Bucur, D.}, \bibinfo{author}{Iacca, G.}, \bibinfo{year}{2016}.
\newblock \bibinfo{title}{Influence maximization in social networks with
  genetic algorithms}, in: \bibinfo{booktitle}{European Conference on the
  Applications of Evolutionary Computation}, \bibinfo{organization}{Springer}.
  pp. \bibinfo{pages}{379--392}.
\bibitem[{Cai et~al.(2017)Cai, He and McAuley}]{cai2017spmc}
\bibinfo{author}{Cai, C.}, \bibinfo{author}{He, R.}, \bibinfo{author}{McAuley,
  J.}, \bibinfo{year}{2017}.
\newblock \bibinfo{title}{Spmc: socially-aware personalized markov chains for
  sparse sequential recommendation}.
\newblock \bibinfo{journal}{arXiv preprint arXiv:1708.04497} .
\bibitem[{Cantador et~al.(2011)Cantador, Brusilovsky and
  Kuflik}]{Cantador:RecSys2011}
\bibinfo{author}{Cantador, I.}, \bibinfo{author}{Brusilovsky, P.},
  \bibinfo{author}{Kuflik, T.}, \bibinfo{year}{2011}.
\newblock \bibinfo{title}{2nd workshop on information heterogeneity and fusion
  in recommender systems (hetrec 2011)}, in: \bibinfo{booktitle}{Proceedings of
  the 5th ACM conference on Recommender systems}, \bibinfo{publisher}{ACM},
  \bibinfo{address}{New York, NY, USA}.
\bibitem[{Carrington et~al.(2005)Carrington, Scott and
  Wasserman}]{carrington2005models}
\bibinfo{author}{Carrington, P.J.}, \bibinfo{author}{Scott, J.},
  \bibinfo{author}{Wasserman, S.}, \bibinfo{year}{2005}.
\newblock \bibinfo{title}{Models and methods in social network analysis}.
  volume~\bibinfo{volume}{28}.
\newblock \bibinfo{publisher}{Cambridge university press}.
\bibitem[{Chen et~al.(2011)Chen, Collins, Cummings, Ke, Liu, Rincon, Sun, Wang,
  Wei and Yuan}]{chen2011influence}
\bibinfo{author}{Chen, W.}, \bibinfo{author}{Collins, A.},
  \bibinfo{author}{Cummings, R.}, \bibinfo{author}{Ke, T.},
  \bibinfo{author}{Liu, Z.}, \bibinfo{author}{Rincon, D.},
  \bibinfo{author}{Sun, X.}, \bibinfo{author}{Wang, Y.}, \bibinfo{author}{Wei,
  W.}, \bibinfo{author}{Yuan, Y.}, \bibinfo{year}{2011}.
\newblock \bibinfo{title}{Influence maximization in social networks when
  negative opinions may emerge and propagate}, in:
  \bibinfo{booktitle}{Proceedings of the 2011 siam international conference on
  data mining}, \bibinfo{organization}{SIAM}. pp. \bibinfo{pages}{379--390}.
\bibitem[{Chen et~al.(2010)Chen, Wang and Wang}]{chen2010scalable}
\bibinfo{author}{Chen, W.}, \bibinfo{author}{Wang, C.}, \bibinfo{author}{Wang,
  Y.}, \bibinfo{year}{2010}.
\newblock \bibinfo{title}{Scalable influence maximization for prevalent viral
  marketing in large-scale social networks}, in:
  \bibinfo{booktitle}{Proceedings of the 16th ACM SIGKDD international
  conference on Knowledge discovery and data mining},
  \bibinfo{organization}{ACM}. pp. \bibinfo{pages}{1029--1038}.
\bibitem[{Chen et~al.(2012)Chen, Chang, Chou, Peng and Lee}]{chen2012exploring}
\bibinfo{author}{Chen, Y.}, \bibinfo{author}{Chang, S.}, \bibinfo{author}{Chou,
  C.}, \bibinfo{author}{Peng, W.}, \bibinfo{author}{Lee, S.},
  \bibinfo{year}{2012}.
\newblock \bibinfo{title}{Exploring community structures for influence
  maximization in social networks}, in: \bibinfo{booktitle}{The 6th SNA-KDD
  Workshop on Social Network Mining and Analysis Held in Conjunction with KDD},
  pp. \bibinfo{pages}{1--6}.
\bibitem[{Chen et~al.(2014)Chen, Zhu, Peng, Lee and Lee}]{chen2014cim}
\bibinfo{author}{Chen, Y.C.}, \bibinfo{author}{Zhu, W.Y.},
  \bibinfo{author}{Peng, W.C.}, \bibinfo{author}{Lee, W.C.},
  \bibinfo{author}{Lee, S.Y.}, \bibinfo{year}{2014}.
\newblock \bibinfo{title}{Cim: Community-based influence maximization in social
  networks}.
\newblock \bibinfo{journal}{ACM Transactions on Intelligent Systems and
  Technology (TIST)} \bibinfo{volume}{5}, \bibinfo{pages}{25}.
\bibitem[{Domingos and Richardson(2001)}]{domingos2001mining}
\bibinfo{author}{Domingos, P.}, \bibinfo{author}{Richardson, M.},
  \bibinfo{year}{2001}.
\newblock \bibinfo{title}{Mining the network value of customers}, in:
  \bibinfo{booktitle}{Proceedings of the seventh ACM SIGKDD international
  conference on Knowledge discovery and data mining},
  \bibinfo{organization}{ACM}. pp. \bibinfo{pages}{57--66}.
\bibitem[{Fan et~al.(2018)Fan, Qiu, Li, Meng, Zhang, Li, Tan and
  Du}]{fan2018octopus}
\bibinfo{author}{Fan, J.}, \bibinfo{author}{Qiu, J.}, \bibinfo{author}{Li, Y.},
  \bibinfo{author}{Meng, Q.}, \bibinfo{author}{Zhang, D.}, \bibinfo{author}{Li,
  G.}, \bibinfo{author}{Tan, K.L.}, \bibinfo{author}{Du, X.},
  \bibinfo{year}{2018}.
\newblock \bibinfo{title}{Octopus: An online topic-aware influence analysis
  system for social networks}, in: \bibinfo{booktitle}{2018 IEEE 34th
  International Conference on Data Engineering (ICDE)},
  \bibinfo{organization}{IEEE}. pp. \bibinfo{pages}{1569--1572}.
\bibitem[{Feige et~al.(2011)Feige, Mirrokni and
  Vondr{\'a}k}]{feige2011maximizing}
\bibinfo{author}{Feige, U.}, \bibinfo{author}{Mirrokni, V.S.},
  \bibinfo{author}{Vondr{\'a}k, J.}, \bibinfo{year}{2011}.
\newblock \bibinfo{title}{Maximizing non-monotone submodular functions}.
\newblock \bibinfo{journal}{SIAM Journal on Computing} \bibinfo{volume}{40},
  \bibinfo{pages}{1133--1153}.
\bibitem[{Galhotra et~al.(2015)Galhotra, Arora, Virinchi and
  Roy}]{galhotra2015asim}
\bibinfo{author}{Galhotra, S.}, \bibinfo{author}{Arora, A.},
  \bibinfo{author}{Virinchi, S.}, \bibinfo{author}{Roy, S.},
  \bibinfo{year}{2015}.
\newblock \bibinfo{title}{Asim: A scalable algorithm for influence maximization
  under the independent cascade model}, in: \bibinfo{booktitle}{Proceedings of
  the 24th International Conference on World Wide Web},
  \bibinfo{organization}{ACM}. pp. \bibinfo{pages}{35--36}.
\bibitem[{Gong et~al.(2016)Gong, Yan, Shen, Ma and Cai}]{gong2016influence}
\bibinfo{author}{Gong, M.}, \bibinfo{author}{Yan, J.}, \bibinfo{author}{Shen,
  B.}, \bibinfo{author}{Ma, L.}, \bibinfo{author}{Cai, Q.},
  \bibinfo{year}{2016}.
\newblock \bibinfo{title}{Influence maximization in social networks based on
  discrete particle swarm optimization}.
\newblock \bibinfo{journal}{Information Sciences} \bibinfo{volume}{367},
  \bibinfo{pages}{600--614}.
\bibitem[{Goyal et~al.(2011)Goyal, Lu and Lakshmanan}]{goyal2011celf}
\bibinfo{author}{Goyal, A.}, \bibinfo{author}{Lu, W.},
  \bibinfo{author}{Lakshmanan, L.V.}, \bibinfo{year}{2011}.
\newblock \bibinfo{title}{Celf++: optimizing the greedy algorithm for influence
  maximization in social networks}, in: \bibinfo{booktitle}{Proceedings of the
  20th international conference companion on World wide web},
  \bibinfo{organization}{ACM}. pp. \bibinfo{pages}{47--48}.
\bibitem[{G{\"u}ney(2017)}]{guney2017optimal}
\bibinfo{author}{G{\"u}ney, E.}, \bibinfo{year}{2017}.
\newblock \bibinfo{title}{On the optimal solution of budgeted influence
  maximization problem in social networks}.
\newblock \bibinfo{journal}{Operational Research} , \bibinfo{pages}{1--15}.
\bibitem[{Han et~al.(2014)Han, Zhuang, He and Shi}]{han2014balanced}
\bibinfo{author}{Han, S.}, \bibinfo{author}{Zhuang, F.}, \bibinfo{author}{He,
  Q.}, \bibinfo{author}{Shi, Z.}, \bibinfo{year}{2014}.
\newblock \bibinfo{title}{Balanced seed selection for budgeted influence
  maximization in social networks}, in: \bibinfo{booktitle}{Pacific-Asia
  Conference on Knowledge Discovery and Data Mining},
  \bibinfo{organization}{Springer}. pp. \bibinfo{pages}{65--77}.
\bibitem[{Hosseini-Pozveh et~al.(2017)Hosseini-Pozveh, Zamanifar and
  Naghsh-Nilchi}]{hosseini2017community}
\bibinfo{author}{Hosseini-Pozveh, M.}, \bibinfo{author}{Zamanifar, K.},
  \bibinfo{author}{Naghsh-Nilchi, A.R.}, \bibinfo{year}{2017}.
\newblock \bibinfo{title}{A community-based approach to identify the most
  influential nodes in social networks}.
\newblock \bibinfo{journal}{Journal of Information Science}
  \bibinfo{volume}{43}, \bibinfo{pages}{204--220}.
\bibitem[{Huang et~al.(2019)Huang, Shen, Meng, Chang and
  He}]{huang2019community}
\bibinfo{author}{Huang, H.}, \bibinfo{author}{Shen, H.}, \bibinfo{author}{Meng,
  Z.}, \bibinfo{author}{Chang, H.}, \bibinfo{author}{He, H.},
  \bibinfo{year}{2019}.
\newblock \bibinfo{title}{Community-based influence maximization for viral
  marketing}.
\newblock \bibinfo{journal}{Applied Intelligence} , \bibinfo{pages}{1--14}.
\bibitem[{Huang et~al.(2020)Huang, Tang, Xiao, Sun and
  Lim}]{huang2020efficient}
\bibinfo{author}{Huang, K.}, \bibinfo{author}{Tang, J.}, \bibinfo{author}{Xiao,
  X.}, \bibinfo{author}{Sun, A.}, \bibinfo{author}{Lim, A.},
  \bibinfo{year}{2020}.
\newblock \bibinfo{title}{Efficient approximation algorithms for adaptive
  target profit maximization}, in: \bibinfo{booktitle}{2020 IEEE 36th
  International Conference on Data Engineering (ICDE)},
  \bibinfo{organization}{IEEE}. pp. \bibinfo{pages}{649--660}.
\bibitem[{Ienco et~al.(2010)Ienco, Bonchi and Castillo}]{ienco2010meme}
\bibinfo{author}{Ienco, D.}, \bibinfo{author}{Bonchi, F.},
  \bibinfo{author}{Castillo, C.}, \bibinfo{year}{2010}.
\newblock \bibinfo{title}{The meme ranking problem: Maximizing microblogging
  virality}, in: \bibinfo{booktitle}{2010 IEEE International Conference on Data
  Mining Workshops}, \bibinfo{organization}{IEEE}. pp.
  \bibinfo{pages}{328--335}.
\bibitem[{Ke et~al.(2018)Ke, Khan and Cong}]{ke2018finding}
\bibinfo{author}{Ke, X.}, \bibinfo{author}{Khan, A.}, \bibinfo{author}{Cong,
  G.}, \bibinfo{year}{2018}.
\newblock \bibinfo{title}{Finding seeds and relevant tags jointly: For targeted
  influence maximization in social networks}, in:
  \bibinfo{booktitle}{Proceedings of the 2018 International Conference on
  Management of Data}, \bibinfo{organization}{ACM}. pp.
  \bibinfo{pages}{1097--1111}.
\bibitem[{Kempe et~al.(2003)Kempe, Kleinberg and Tardos}]{kempe2003maximizing}
\bibinfo{author}{Kempe, D.}, \bibinfo{author}{Kleinberg, J.},
  \bibinfo{author}{Tardos, {\'E}.}, \bibinfo{year}{2003}.
\newblock \bibinfo{title}{Maximizing the spread of influence through a social
  network}, in: \bibinfo{booktitle}{Proceedings of the ninth ACM SIGKDD
  international conference on Knowledge discovery and data mining},
  \bibinfo{organization}{ACM}. pp. \bibinfo{pages}{137--146}.
\bibitem[{Leskovec et~al.(2007)Leskovec, Krause, Guestrin, Faloutsos,
  VanBriesen and Glance}]{leskovec2007cost}
\bibinfo{author}{Leskovec, J.}, \bibinfo{author}{Krause, A.},
  \bibinfo{author}{Guestrin, C.}, \bibinfo{author}{Faloutsos, C.},
  \bibinfo{author}{VanBriesen, J.}, \bibinfo{author}{Glance, N.},
  \bibinfo{year}{2007}.
\newblock \bibinfo{title}{Cost-effective outbreak detection in networks}, in:
  \bibinfo{booktitle}{Proceedings of the 13th ACM SIGKDD international
  conference on Knowledge discovery and data mining},
  \bibinfo{organization}{ACM}. pp. \bibinfo{pages}{420--429}.
\bibitem[{Li et~al.(2018a)Li, Cheng, Su and Sun}]{li2018community}
\bibinfo{author}{Li, X.}, \bibinfo{author}{Cheng, X.}, \bibinfo{author}{Su,
  S.}, \bibinfo{author}{Sun, C.}, \bibinfo{year}{2018}a.
\newblock \bibinfo{title}{Community-based seeds selection algorithm for
  location aware influence maximization}.
\newblock \bibinfo{journal}{Neurocomputing} \bibinfo{volume}{275},
  \bibinfo{pages}{1601--1613}.
\bibitem[{Li et~al.(2018b)Li, Fan, Wang and Tan}]{li2018influence}
\bibinfo{author}{Li, Y.}, \bibinfo{author}{Fan, J.}, \bibinfo{author}{Wang,
  Y.}, \bibinfo{author}{Tan, K.L.}, \bibinfo{year}{2018}b.
\newblock \bibinfo{title}{Influence maximization on social graphs: A survey}.
\newblock \bibinfo{journal}{IEEE Transactions on Knowledge and Data
  Engineering} \bibinfo{volume}{30}, \bibinfo{pages}{1852--1872}.
\bibitem[{Li et~al.(2015)Li, Zhang and Tan}]{li2015real}
\bibinfo{author}{Li, Y.}, \bibinfo{author}{Zhang, D.}, \bibinfo{author}{Tan,
  K.L.}, \bibinfo{year}{2015}.
\newblock \bibinfo{title}{Real-time targeted influence maximization for online
  advertisements}.
\newblock \bibinfo{journal}{Proceedings of the VLDB Endowment}
  \bibinfo{volume}{8}, \bibinfo{pages}{1070--1081}.
\bibitem[{Liu et~al.(2020)Liu, Li, Wang, Fang, Dong and Wu}]{liu2020profit}
\bibinfo{author}{Liu, B.}, \bibinfo{author}{Li, X.}, \bibinfo{author}{Wang,
  H.}, \bibinfo{author}{Fang, Q.}, \bibinfo{author}{Dong, J.},
  \bibinfo{author}{Wu, W.}, \bibinfo{year}{2020}.
\newblock \bibinfo{title}{Profit maximization problem with coupons in social
  networks}.
\newblock \bibinfo{journal}{Theoretical Computer Science}
  \bibinfo{volume}{803}, \bibinfo{pages}{22--35}.
\bibitem[{Narayanam and Narahari(2011)}]{narayanam2011shapley}
\bibinfo{author}{Narayanam, R.}, \bibinfo{author}{Narahari, Y.},
  \bibinfo{year}{2011}.
\newblock \bibinfo{title}{A shapley value-based approach to discover
  influential nodes in social networks}.
\newblock \bibinfo{journal}{IEEE Transactions on Automation Science and
  Engineering} \bibinfo{volume}{8}, \bibinfo{pages}{130--147}.
\bibitem[{Nguyen and Zheng(2013)}]{nguyen2013budgeted}
\bibinfo{author}{Nguyen, H.}, \bibinfo{author}{Zheng, R.},
  \bibinfo{year}{2013}.
\newblock \bibinfo{title}{On budgeted influence maximization in social
  networks}.
\newblock \bibinfo{journal}{IEEE Journal on Selected Areas in Communications}
  \bibinfo{volume}{31}, \bibinfo{pages}{1084--1094}.
\bibitem[{Peng et~al.(2018)Peng, Zhou, Cao, Yu, Niu and
  Jia}]{peng2018influence}
\bibinfo{author}{Peng, S.}, \bibinfo{author}{Zhou, Y.}, \bibinfo{author}{Cao,
  L.}, \bibinfo{author}{Yu, S.}, \bibinfo{author}{Niu, J.},
  \bibinfo{author}{Jia, W.}, \bibinfo{year}{2018}.
\newblock \bibinfo{title}{Influence analysis in social networks: A survey}.
\newblock \bibinfo{journal}{Journal of Network and Computer Applications}
  \bibinfo{volume}{106}, \bibinfo{pages}{17--32}.
\bibitem[{Shang et~al.(2017)Shang, Zhou, Li, Liu and Wu}]{shang2017cofim}
\bibinfo{author}{Shang, J.}, \bibinfo{author}{Zhou, S.}, \bibinfo{author}{Li,
  X.}, \bibinfo{author}{Liu, L.}, \bibinfo{author}{Wu, H.},
  \bibinfo{year}{2017}.
\newblock \bibinfo{title}{Cofim: A community-based framework for influence
  maximization on large-scale networks}.
\newblock \bibinfo{journal}{Knowledge-Based Systems} \bibinfo{volume}{117},
  \bibinfo{pages}{88--100}.
\bibitem[{Singh et~al.(2019)Singh, Kumar, Singh and Biswas}]{singh2019c2im}
\bibinfo{author}{Singh, S.S.}, \bibinfo{author}{Kumar, A.},
  \bibinfo{author}{Singh, K.}, \bibinfo{author}{Biswas, B.},
  \bibinfo{year}{2019}.
\newblock \bibinfo{title}{C2im: Community based context-aware influence
  maximization in social networks}.
\newblock \bibinfo{journal}{Physica A: Statistical Mechanics and its
  Applications} \bibinfo{volume}{514}, \bibinfo{pages}{796--818}.
\bibitem[{Tang et~al.(2017)Tang, Tang and Yuan}]{tang2017profit}
\bibinfo{author}{Tang, J.}, \bibinfo{author}{Tang, X.}, \bibinfo{author}{Yuan,
  J.}, \bibinfo{year}{2017}.
\newblock \bibinfo{title}{Profit maximization for viral marketing in online
  social networks: Algorithms and analysis}.
\newblock \bibinfo{journal}{IEEE Transactions on Knowledge and Data
  Engineering} \bibinfo{volume}{30}, \bibinfo{pages}{1095--1108}.
\bibitem[{Tang et~al.(2015)Tang, Shi and Xiao}]{tang2015influence}
\bibinfo{author}{Tang, Y.}, \bibinfo{author}{Shi, Y.}, \bibinfo{author}{Xiao,
  X.}, \bibinfo{year}{2015}.
\newblock \bibinfo{title}{Influence maximization in near-linear time: A
  martingale approach}, in: \bibinfo{booktitle}{Proceedings of the 2015 ACM
  SIGMOD International Conference on Management of Data},
  \bibinfo{organization}{ACM}. pp. \bibinfo{pages}{1539--1554}.
\bibitem[{Tang et~al.(2014)Tang, Xiao and Shi}]{tang2014influence}
\bibinfo{author}{Tang, Y.}, \bibinfo{author}{Xiao, X.}, \bibinfo{author}{Shi,
  Y.}, \bibinfo{year}{2014}.
\newblock \bibinfo{title}{Influence maximization: Near-optimal time complexity
  meets practical efficiency}, in: \bibinfo{booktitle}{Proceedings of the 2014
  ACM SIGMOD international conference on Management of data},
  \bibinfo{organization}{ACM}. pp. \bibinfo{pages}{75--86}.
\bibitem[{Tong et~al.(2018)Tong, Wu and Du}]{tong2018coupon}
\bibinfo{author}{Tong, G.}, \bibinfo{author}{Wu, W.}, \bibinfo{author}{Du,
  D.Z.}, \bibinfo{year}{2018}.
\newblock \bibinfo{title}{Coupon advertising in online social systems:
  Algorithms and sampling techniques}.
\newblock \bibinfo{journal}{arXiv preprint arXiv:1802.06946} .
\bibitem[{Wang et~al.(2012)Wang, Chen and Wang}]{wang2012scalable}
\bibinfo{author}{Wang, C.}, \bibinfo{author}{Chen, W.}, \bibinfo{author}{Wang,
  Y.}, \bibinfo{year}{2012}.
\newblock \bibinfo{title}{Scalable influence maximization for independent
  cascade model in large-scale social networks}.
\newblock \bibinfo{journal}{Data Mining and Knowledge Discovery}
  \bibinfo{volume}{25}, \bibinfo{pages}{545--576}.
\bibitem[{Wang et~al.(2010)Wang, Cong, Song and Xie}]{wang2010community}
\bibinfo{author}{Wang, Y.}, \bibinfo{author}{Cong, G.}, \bibinfo{author}{Song,
  G.}, \bibinfo{author}{Xie, K.}, \bibinfo{year}{2010}.
\newblock \bibinfo{title}{Community-based greedy algorithm for mining top-k
  influential nodes in mobile social networks}, in:
  \bibinfo{booktitle}{Proceedings of the 16th ACM SIGKDD international
  conference on Knowledge discovery and data mining},
  \bibinfo{organization}{ACM}. pp. \bibinfo{pages}{1039--1048}.
\bibitem[{Zhang et~al.(2013)Zhang, Wang and Vassileva}]{zhang2013socconnect}
\bibinfo{author}{Zhang, J.}, \bibinfo{author}{Wang, Y.},
  \bibinfo{author}{Vassileva, J.}, \bibinfo{year}{2013}.
\newblock \bibinfo{title}{Socconnect: A personalized social network aggregator
  and recommender}.
\newblock \bibinfo{journal}{Information Processing \& Management}
  \bibinfo{volume}{49}, \bibinfo{pages}{721--737}.
\bibitem[{Zhang et~al.(2017)Zhang, Du and Feldman}]{zhang2017maximizing}
\bibinfo{author}{Zhang, K.}, \bibinfo{author}{Du, H.},
  \bibinfo{author}{Feldman, M.W.}, \bibinfo{year}{2017}.
\newblock \bibinfo{title}{Maximizing influence in a social network: Improved
  results using a genetic algorithm}.
\newblock \bibinfo{journal}{Physica A: Statistical Mechanics and its
  Applications} \bibinfo{volume}{478}, \bibinfo{pages}{20--30}.
\bibitem[{Zhao et~al.(2015)Zhao, McAuley and King}]{zhao2015improving}
\bibinfo{author}{Zhao, T.}, \bibinfo{author}{McAuley, J.},
  \bibinfo{author}{King, I.}, \bibinfo{year}{2015}.
\newblock \bibinfo{title}{Improving latent factor models via personalized
  feature projection for one class recommendation}, in:
  \bibinfo{booktitle}{Proceedings of the 24th ACM international on conference
  on information and knowledge management}, \bibinfo{organization}{ACM}. pp.
  \bibinfo{pages}{821--830}.
\bibitem[{Zhou et~al.(2019)Zhou, Fan, Wang, Wang and Li}]{zhou2019cost}
\bibinfo{author}{Zhou, J.}, \bibinfo{author}{Fan, J.}, \bibinfo{author}{Wang,
  J.}, \bibinfo{author}{Wang, X.}, \bibinfo{author}{Li, L.},
  \bibinfo{year}{2019}.
\newblock \bibinfo{title}{Cost-efficient viral marketing in online social
  networks}.
\newblock \bibinfo{journal}{World Wide Web} \bibinfo{volume}{22},
  \bibinfo{pages}{2355--2378}.
\bibitem[{Zhu et~al.(2017)Zhu, Li, Yan, Wu and Bi}]{zhu2017maximizing}
\bibinfo{author}{Zhu, Y.}, \bibinfo{author}{Li, D.}, \bibinfo{author}{Yan, R.},
  \bibinfo{author}{Wu, W.}, \bibinfo{author}{Bi, Y.}, \bibinfo{year}{2017}.
\newblock \bibinfo{title}{Maximizing the influence and profit in social
  networks}.
\newblock \bibinfo{journal}{IEEE Transactions on Computational Social Systems}
  \bibinfo{volume}{4}, \bibinfo{pages}{54--64}.

\end{thebibliography}

\end{document}